%% file: main.tex
\documentclass[11pt,a4paper]{article}
\pdfoutput=1 

\usepackage{jheppub}

\usepackage[utf8]{inputenc}
\usepackage[T1]{fontenc}
\usepackage{lmodern}
\usepackage{fontawesome}
\usepackage{hyperref}
\usepackage{tikz}
\usepackage{subcaption}
\usepackage{mathtools}
\usepackage{xcolor}
\usepackage{amsmath,amssymb}
\usepackage{booktabs}
\usepackage{bbding}
\usepackage{graphicx} 
\usepackage{tikz-feynman}
\usepackage{bm}
\usepackage{multirow}

\newcommand{\smefit}{{\sc \small SMEFiT} }
\newcommand{\flavio}{{\sc \small flavio} }
\newcommand{\OO}{\mathcal{O}}

\title{Probing the flavour-blind SMEFT: EFT validity and the interplay of energy scales}
\author{Luca Mantani and}
\author{Veronica Sanz}

\affiliation{Instituto de F\'isica Corpuscular (IFIC), Universidad de Valencia-CSIC, E-46980 Valencia, Spain}

\emailAdd{luca.mantani@uv.es}
\emailAdd{veronica.sanz@uv.es}

\abstract{
The Standard Model Effective Field Theory (SMEFT) offers a systematic approach to study potential deviations from the Standard Model (SM) through higher-dimensional operators that encapsulate new physics effects. In this work, we analyze flavour-blind SMEFT contributions to flavour observables and assess their interplay with high-energy measurements from LEP and LHC. We perform global fits combining LEP precision data, flavour observables from rare B-meson decays, and LHC diboson measurements, revealing how the inclusion of different datasets breaks parameter degeneracies and enhances the sensitivity to SMEFT coefficients. Our study demonstrates that low-energy flavour observables provide reliable constraints even in flavour-blind scenarios, while high-energy measurements can be subject to EFT validity concerns due to kinematic growth. We investigate the impact of renormalization group evolution (RGE) and operator mixing across energy scales, highlighting the complementary nature of low- and high-energy datasets. The results emphasize the importance of flavour observables as robust probes of new physics and underline the necessity of global fits to avoid potential biases from limited datasets. Finally, we discuss the implications of our findings for the interpretation of global SMEFT analyses based on high-energy collider data, comparing UV models that contribute to SMEFT at tree- and loop-level.

}

\begin{document}
\maketitle


\input{sec-intro}

\input{sec-theory-settings}

\input{sec-LEP-data}

\input{sec-results}

\input{sec-UV-completions}

\input{sec-conclusions}

\section*{Aknowledgments}
LM and VS would like to thank Eleni Vryonidou for the careful reading of the manuscript and her useful criticisms. LM thanks Marco Fedele  for useful discussions. 
LM acknowledges support from the European Union under the MSCA fellowship (Grant agreement N. 101149078) {\it Advancing global SMEFT fits in the LHC precision era (EFT4ward)}.
The research of VS is supported by the Generalitat
Valenciana PROMETEO/2021/083 and Proyecto Consolidacion CNS2022-135688 from the AEI. The research of LM and VS is supported by  the Ministerio de Ciencia e
Innovacion project PID2023-148162NB-C21, and the {\it Severo Ochoa} project CEX2023-001292-S funded by MCIU/AEI. 

\newpage
\appendix
\input{app-pca}
\input{app-bounds}

\bibliographystyle{JHEP}
\bibliography{references.bib}
\end{document}

%% file: sec-intro.tex
\section{Introduction}

The Standard Model (SM) of particle physics has been remarkably successful in explaining a wide range of experimental observations, especially in the realm of low-energy phenomena. However, it is widely believed that the SM is not the ultimate theory, but rather an effective description valid up to a certain energy scale, beyond which new physics might emerge. The Standard Model Effective Field Theory (SMEFT)~\cite{Grzadkowski_2010,Brivio_2019,isidori2023standardmodeleffectivefield} provides a framework to study potential heavy new physics effects in a systematic and model-independent way, by encapsulating deviations from the SM through higher-dimensional operators. In recent years, the development of global interpretation frameworks has become a key priority, aiming to fully harness the potential of the SMEFT in uncovering deviations from the SM~\cite{de_Blas_2020,Brivio_2022,Ellis_2021,Giani_2023}.

Low-energy observables, particularly in the flavour sector, have proven to be sensitive probes for testing the structure of the SM and searching for signs of new physics~\cite{Cirigliano:2016nyn, Cirigliano:2023nol, Garosi:2023yxg}. These observables allow us to explore the intricate flavour structure of the SM with high precision and have been instrumental in providing stringent constraints on possible extensions of the theory. Although flavour observables are often most powerful in scenarios involving new physics with non-trivial flavour structures, they are also effective in studying flavour-blind extensions, such as axions, axion-like particles (ALPs)~\cite{Izaguirre:2016dfi, Bauer:2021mvw, MartinCamalich:2020dfe, Gavela:2019wzg, Albrecht:2019zul}, or other models where flavour symmetries are preserved.

There is a substantial body of work exploring the addition of non-trivial flavour structures within the SMEFT framework~\cite{Allwicher:2023shc,Aoude:2020dwv, Bruggisser:2021duo, Bruggisser:2022rhb, Bartocci:2023nvp, Bellafronte:2023amz}, where new physics is expected to interact differently with various generations of quarks and leptons. These studies introduce rich and complex flavour dynamics, offering promising avenues to explain existing tensions or anomalies in flavour physics experiments. However, this is not the focus of our paper. Instead, we adopt a flavour-blind approach to new physics, assuming that any beyond-the-Standard-Model interactions respect flavour symmetries and can be described using the SMEFT framework without introducing flavour-dependent operators.

Typically, flavour-blind SMEFT searches exploit the kinematic growth of certain observables at high-energy colliders, where the effects of new physics are enhanced~\cite{Farina:2016rws, Dror:2015nkp, Maltoni:2019aot}. However, this kinematic growth also signals that the effective field theory (EFT) expansion may be breaking down, as higher-dimensional operators become more relevant at such energy scales. Therefore, it is crucial to examine other types of constraints that do not rely on kinematic enhancements, such as low-energy measurements. Although these low-energy observables might seem nominally less sensitive, they provide valuable and robust tests of new physics without the uncertainties associated with the breakdown of the EFT expansion.

In this paper, we investigate how flavour observables are affected by such flavour-blind new physics, taking into account the renormalization group evolution (RGE) and operator mixing that occurs between different energy scales~\cite{Jenkins_2013,Jenkins:2013sda,Jenkins_2014,aoude2023renormalisationgroupeffectssmeft, Maltoni:2024dpn, Bartocci:2024fmm, terHoeve:2025gey}. By studying these effects, we aim to show how flavour observables can still serve as a powerful tool to probe beyond the SM, even in flavour-blind contexts.

This paper is organized as follows. In Sec.~\ref{sec:theory} we describe the SMEFT operators relevant to our analysis, with a particular emphasis on trilinear gauge boson couplings and their connection to high-energy and low-energy observables. In Sec.~\ref{sec:legacy}, we review the legacy LEP data and its role in constraining SMEFT coefficients, highlighting key observables such as Z pole measurements, Bhabha scattering, and WW diboson production. In Sec.~\ref{sec:globalfits}, we present global fits that combine LEP, flavour, and LHC data. In particular,  Sec.~\ref{sec:flavour_observables} focuses on flavour observables and their sensitivity to flavour-blind SMEFT contributions. In Sec.~\ref{sec:uv}, we discuss the implications of the UV origins of SMEFT coefficients for the validity of the EFT description in different datasets, emphasizing the challenges posed by high-energy regimes and the robustness of low-energy observables. Finally, Sec.~\ref{sec:concls} summarizes our findings and outlines prospects for future research.

%% file: sec-theory-settings.tex
\section{Theoretical framework}
\label{sec:theory}
\input{operators_table}

In this section, we establish the theoretical framework underlying our analysis. Specifically, we consider an extension of the SM within the SMEFT formalism. The key assumption is that potential new physics (NP) degrees of freedom are heavier than the energy scales currently accessible in experiments. Under this premise, their effects can be systematically encoded in higher-dimensional operators, modifying interactions in a model-independent manner. At leading order, the dominant corrections to collider observables arise from dimension-six operators~\cite{Grzadkowski_2010}.

The SMEFT Lagrangian extends the SM as follows:
\begin{equation}
    \mathcal{L} = \mathcal{L}_{SM} + \sum_{i}^{n_{op}} \frac{c_i}{\Lambda^2} \mathcal{O}_i \, ,
\end{equation}
where $\Lambda$ represents the characteristic NP scale suppressing the effects of higher-dimensional operators, and $n_{op}$ denotes the number of independent operators, which depends on the assumed flavour symmetries. In the most general, flavour-dependent case, up to $n_{op} = 2499$ operators are introduced. However, in our analysis, we adopt the opposite limit and assume a flavour-blind NP scenario, leading to a minimal operator basis with $n_{op} = 59$.

Approaching the detection of NP agnostically from a bottom-up perspective allows for a reduction in the number of relevant operators by initially focusing on subsets of observables that are sensitive to only a limited portion of the full operator basis. In this approach, a targeted subset of operators can be defined, allowing for the investigation of potential constraints on them and assessing whether observed deviations in the data can be consistently interpreted within this framework. However, we emphasize that this approach is inherently limited and may not always be well-defined. The SMEFT framework modifies interactions in a way that induces correlations among different observables, and only a truly global analysis can reliably account for these effects~\cite{Ellis:2020unq, Celada:2024mcf, Ethier:2021bye, Kassabov:2023hbm}. This consideration becomes particularly crucial when including RGE effects or higher-order loop corrections~\cite{Bartocci:2024fmm, terHoeve:2025gey}, as operator mixing across different energy scales complicates the isolation of specific operator subsets.

Our analysis includes all the operators listed in Table~\ref{tab:ops}, which account for the relevant contributions to EWPOs and diboson production at LEP. The table reports the limits on the Wilson coefficients obtained from the global fit in Ref.~\cite{Celada:2024mcf}. For operators that were not assumed to be flavor-universal, we provide the most stringent bound among the different flavours. This selection is driven by our goal of examining the interplay between LEP data and other sectors, particularly flavour physics. Consequently, we will focus on these operators in the following discussion.
These operators also contribute to LHC observables, notably affecting diboson production. They modify the interactions between fermions and electroweak bosons, both right- and left-handed couplings, as well as the triple gauge couplings (TGCs). The latter is of particular interest as TGCs are expected to be modified in several BSM extensions such as composite Higgs models, SUSY and extended gauge sectors (presence of $Z^\prime$ and $W^\prime$), among others.

These operators can also be constrained using flavour observables, particularly through their effects on rare $b$ decays. Since these processes occur at energies significantly below the electroweak scale, a natural framework to describe interactions at the $b$-quark mass scale is the Weak Effective Theory (WET). In WET, the top quark, Higgs boson, and electroweak gauge bosons are integrated out, and their mediated interactions are encoded in higher-dimensional operators. 

In recent years, significant efforts have been devoted to establishing precise matching relations between SMEFT and WET operators~\cite{Hurth:2019ula, Aoude:2020dwv}. Notably, Ref.~\cite{Hurth:2019ula} provides one-loop matching calculations for the SMEFT operators considered in this study, focusing on those relevant for flavour-changing neutral currents (FCNCs).
In particular, we consider the WET Lagrangian:

\begin{equation}
    \mathcal{L}_{|\Delta B|=|\Delta S|=1}=\frac{4 G_F}{\sqrt{2}} V_{t s}^* V_{t b}\left(C_7 \OO_7+C_9 \OO_9+C_{10} \OO_{10}\right)+\text { h.c. } \, ,
\end{equation}
where the Wilson coefficients are $C_i = C_i^{SM} + \Delta C_i$ and the WET operators are given by
\begin{align}
\OO_7 & =\frac{e}{(4 \pi)^2} m_b\left(\bar{s}_L \sigma_{\alpha \beta} b_R\right) F^{\alpha \beta} \\
\OO_9 & =\frac{e^2}{(4 \pi)^2}\left(\bar{s}_L \gamma_\alpha b_L\right)\left(\bar{\ell} \gamma^\alpha \ell\right) \\
\OO_{10} & =\frac{e^2}{(4 \pi)^2}\left(\bar{s}_L \gamma_\alpha b_L\right)\left(\bar{\ell} \gamma^\alpha \gamma_5 \ell\right) \, .
\end{align}

The matching expressions are provided at the electroweak scale $m_W$, assuming SMEFT Wilson coefficients defined at a generic high-energy scale $\Lambda$. We employ the relations from Ref.~\cite{Hurth:2019ula} to extend our SMEFT operator dependence to flavour observables, with the objective of incorporating data that is sensitive to the three WET operators discussed above.

When considering observables at substantially different energy scales, as it is the case for $b$ physics, EWPOs and diboson production at the LHC, it becomes paramount to include the RGE running effects of the Wilson coefficients. These dictate the energy dependence of the Wilson coefficients and their mixing. Throughout this study we assess their impact by leveraging the \smefit framework functionalities~\cite{terHoeve:2025gey}.

Throughout this study, we consider only linear corrections in the EFT at dimension six, neglecting quadratic contributions. However, in Appendix~\ref{app:bounds}, we also report results obtained when including quadratic corrections for completeness. Theoretical calculations are performed within the $\{m_W, m_h, m_Z, G_F\}$ electroweak input scheme, commonly referred to as the $m_W$ input scheme. Furthermore, we define the Wilson coefficients at a reference scale $\mu_0$, chosen as either the NP scale $\Lambda$ or the $Z$ boson mass $m_Z$, depending on the context. The numerical values used for the input parameters are:  
\begin{align*}
    m_W &= 80.387\,{\rm GeV}, \qquad \;\;
    m_h = 125\,{\rm GeV}, \\
    m_Z &= 91.1876\,{\rm GeV}, \qquad
    G_F = 1.1663787 \times 10^{-5}\,{\rm GeV}^{-2}, \\
    \Lambda &= 1\,{\rm TeV}.
\end{align*}

%% file: operators_table.tex
\begin{table}[t!]
  \begin{center}
    \renewcommand{\arraystretch}{1.4}
    {\small
    \begin{tabular}{ccccc}
      \toprule
      Operator $\qquad$ & Coefficient & Definition & 95\,\% CL $\mathcal{O}(1/\Lambda^2)$ & 95\,\% CL $\mathcal{O}(1/\Lambda^4)$\\
      \hline\bottomrule
    \multicolumn{4}{c}{two-fermion operators}\\
    \toprule
    $\OO_{\varphi u}$ & $c_{\varphi u}$ & $i\big(\varphi^\dagger \overset{\leftrightarrow}{D}_\mu\varphi\big)\big(\bar{u}\gamma^\mu u\big)$ & [-0.375, 0.461] & [-0.168, 0.177] \\\hline
    $\OO_{\varphi d}$ & $c_{\varphi d}$ & $i\big(\varphi^\dagger \overset{\leftrightarrow}{D}_\mu\varphi\big)\big(\bar{d}\gamma^\mu d\big)$ & [-1.038, 0.030] & [-0.303, 0.143] \\\hline
    $\OO_{\varphi q}^{(1)}$ & $c_{\varphi q}^{(1)}$ & $i\big(\varphi^\dagger \overset{\leftrightarrow}{D}_\mu\,\varphi\big)
 \big(\bar{q}\,\gamma^\mu\,q\big)$ & - & - \\\hline
  $\OO_{\varphi q}^{(-)}$ & $c_{\varphi q}^{(-)}$ & $c_{\varphi q}^{(1)} - c_{\varphi q}^{(3)}$ & [-0.193, 0.269] & [-0.056, 0.239] \\\hline
    $\OO_{\varphi q}^{(3)}$ & $c_{\varphi q}^{(3)}$  & $i\big(\varphi^\dagger\overset{\leftrightarrow}{D}_\mu\,\tau_{ I}\varphi\big)
 \big(\bar{q}\,\gamma^\mu\,\tau^{ I}q\big)$ & [-0.147, -0.002] & [-0.166, -0.010] \\ \hline
    $\OO_{\varphi e}$ & $c_{\varphi e}$ & $i\big(\varphi^\dagger \overset{\leftrightarrow}{D}_\mu\varphi\big)\big(\bar{e}\gamma^\mu e\big)$ & [-0.583, 0.527] & [-0.254, 0.248] \\\hline
    $\OO_{\varphi l}^{(1)}$ & $c_{\varphi l}^{(1)}$ & $i\big(\varphi^\dagger \overset{\leftrightarrow}{D}_\mu\varphi\big)\big(\bar{l}\gamma^\mu l\big)$ & [-0.276, 0.273] & [-0.133, 0.150]\\\hline
    $\OO_{\varphi l}^{(3)}$ & $c_{\varphi l}^{(3)}$ & $i\big(\varphi^\dagger \overset{\leftrightarrow}{D}_\mu \tau_I\varphi\big)\big(\bar{l}\gamma^\mu \tau^I l\big)$ & [-0.136, 0.064] & [-0.170, 0.027]\\\hline
 \bottomrule
 \multicolumn{4}{c}{bosonic operators}\\
 \toprule
$\OO_W$ &  $c_{W}$ & $\varepsilon_{IJK} W^I_{\mu\nu}W^{J,\nu\rho} W^{K,\mu}_\rho$,  & [-0.565, 0.609] & [-0.156, 0.230] \\\hline
$\OO_{\varphi WB}$ &   $c_{\varphi W B}$ & $(\varphi^\dagger \tau_I \varphi) B^{\mu\nu} W^I_{\mu\nu}$ & [-0.525, 0.504] & [-0.190, 0.263] \\\hline
$\OO_{\varphi D}$ &   $c_{\varphi D}$ & $(\varphi^\dagger D^\mu \varphi)^\dagger (\varphi^\dagger D_\mu \varphi)$ & [-1.063, 1.149] & [-0.513, 0.483] \\\hline
  \bottomrule
  \multicolumn{4}{c}{four-fermion operator}\\
  \toprule
  $\OO_{ll}$ &  $c_{ll}$ & $\big(\bar{l}\gamma_\mu l\big)\big(\bar{l}\gamma^\mu l\big)$ & [-0.112, 0.100] & [-0.066, 0.149]   \\
  \hline\bottomrule
\end{tabular}
}
\end{center}
  \caption{Definition of the dimension-six SMEFT operators.
    The bounds on the Wilson coefficients assume a scale of $\Lambda=1$ TeV and are taken from the global fit of Ref.~\cite{Celada:2024mcf} both at order $\mathcal{O}(\Lambda^{-2})$ and $\mathcal{O}(\Lambda^{-4})$. \label{tab:ops}}
\end{table}

%% file: sec-LEP-data.tex
\section{Legacy LEP data}
\label{sec:legacy}

LEP provided a wealth of high-precision experimental data that remains instrumental for probing the SM and constraining new physics frameworks such as the SMEFT. The diverse range of measurements obtained at LEP includes precision observables at the Z pole, Bhabha scattering data, and WW diboson production, which collectively offer stringent tests of the SM and sensitivity to subtle deviations induced by flavour-blind SMEFT operators. We further complement this data with the measurement of $\alpha_{EW}$ at the Z-pole~\cite{10.1093/ptep/ptac097}.

Key observables at the Z resonance include the total and partial decay widths of the Z boson, forward-backward asymmetries, and polarization asymmetries. These measurements achieve sub-percent-level precision, offering a detailed probe of electroweak interactions. Flavour-blind SMEFT operators, which respect the SM's flavour symmetries, can modify these observables by introducing shifts in the electroweak couplings or altering loop-level corrections, providing sensitivity to such effects~\cite{Brivio:2017bnu}.

In particular, dimension-six operators modify the couplings of the $Z$ boson to fermions, directly affecting all the above observables. In the $m_W$ scheme, the extraction of input parameters is shifted as follows:
\begin{align}
    \delta G_F &= \frac{1}{2 \hat{G}_F} \big( 2 c_{\varphi \ell}^{(3)} -  c_{\ell\ell}  \big) \, , \nonumber\\
    \frac{\delta m_Z^2}{\hat{m}_Z^2} &= \frac{1}{2\sqrt{2}\hat{G}_F}c_{\varphi D} + \frac{\sqrt{2}}{\hat{G}_F}\frac{\hat{m}_W}{\hat{m}_Z}\sqrt{1-\frac{\hat{m}_W^2}{\hat{m}_Z^2}}c_{\varphi WB} \, , \label{eq:input_EW_shifts}\\
    \frac{\delta m_W^2}{\hat{m}_W^2} &= 0 \, . \nonumber
\end{align}
In this notation, the scale of new physics $\Lambda$ is absorbed in the Wilson coefficients.

The $Z$ boson couplings are then shifted both due to these indirect input shifts and direct operator insertions. Following Ref.~\cite{Brivio:2017bnu}, we cast the generic axial and vector coupling of fermions to the $Z$ boson as:
\begin{align}
    g_{V, A}^x = \bar{g}_{V, A}^x + \delta g_{V, A}^x \, , \qquad x=\{\ell_i, u_i, d_i, \nu_i\} \, ,
\end{align}
where SM couplings are given by:
\begin{equation}
    \bar{g}_V^x = \frac{T_3^x}{2} - Q^x s_{\hat{\theta}}^2  \, ,\qquad \bar{g}_{A}^{x} = \frac{T_3^x}{2} \, ,
\end{equation}
with charge and weak isospin values:
\begin{equation}
    Q^x = \{-1, 2/3, -1/3, 0\} \, ,\quad T_3^x = \{-1/2, 1/2, -1/2, 1/2\} \, ,\quad s_{\hat{\theta}}^2=1-\frac{\hat{m}_W^2}{\hat{m}_Z^2} \, .
\end{equation}
The anomalous couplings can be further decomposed as:
\begin{align}
    \delta g_V^x &= \delta\bar{g}_Z\,\bar{g}_V^x + Q^x\delta s_\theta^2 + \Delta_V^x, \\
    \delta g_A^x &= \delta\bar{g}_Z\,\bar{g}_A + \Delta_A^x \, ,
\end{align}
where:
\begin{align}
    \delta \bar{g}_Z &= -\frac{1}{\sqrt{2}}\delta G_F - \frac{1}{2}\frac{\delta m_Z^2}{\hat{m}_Z^2} + \frac{s_{\hat{\theta}}c_{\hat{\theta}}}{\sqrt{2}\hat{G}_F}c_{\varphi WB}\\
    &= -\frac{1}{4\sqrt{2} \hat{G}_F}\big(c_{\varphi D} + 4 c_{\varphi \ell}^{(3)} - 2c_{\ell\ell}\big) \, ,\\
    \delta s_{\theta}^2 &= \frac{1}{2\sqrt{2}\hat{G}_F}\frac{\hat{m}_W^2}{\hat{m}_Z^2}c_{\varphi D} + \frac{1}{\sqrt{2}\hat{G}_F}\frac{\hat{m}_W}{\hat{m}_Z}\sqrt{1-\frac{\hat{m}_W^2}{\hat{m}_Z^2}}c_{\varphi WB} \, .
\end{align}
Specifically, the direct contributions to the anomalous couplings in terms of the Wilson coefficients in Tab.~\ref{tab:ops} are given by:
\begin{align}
    \Delta^{\ell}_V &= -\frac{1}{4\sqrt{2} \hat{G}_F}\big(c_{\varphi\ell}^{(1)}+c_{\varphi\ell}^{(3)}+c_{\varphi e}\big) \, , & \Delta^{\ell}_A &= -\frac{1}{4\sqrt{2} \hat{G}_F}\big(c_{\varphi \ell}^{(1)}+c_{\varphi\ell}^{(3)}-c_{\varphi e}\big),\\
    \Delta^{\nu}_V &= -\frac{1}{4\sqrt{2} \hat{G}_F}\big(c_{\varphi\ell}^{(1)}-c_{\varphi\ell}^{(3)} \big) \, , & \Delta^{\nu}_A &= -\frac{1}{4\sqrt{2} \hat{G}_F}\big(c_{\varphi \ell}^{(1)}-c_{\varphi\ell}^{(3)}\big) \, ,\\
    \Delta^{u}_V &= -\frac{1}{4\sqrt{2} \hat{G}_F}\big(c_{\varphi q}^{(1)}-c_{\varphi q}^{(3)}+c_{\varphi u}\big) \, , & \Delta^{u}_A &= -\frac{1}{4\sqrt{2} \hat{G}_F}\big(c_{\varphi q}^{(1)}-c_{\varphi q}^{(3)}-c_{\varphi u}\big),\\
    \Delta^{d}_V &= -\frac{1}{4\sqrt{2} \hat{G}_F}\big(c_{\varphi q}^{(1)}+c_{\varphi q}^{(3)}+c_{\varphi d}\big) \, , & \Delta^{d}_A &= -\frac{1}{4\sqrt{2} \hat{G}_F}\big(c_{\varphi q}^{(1)}+c_{\varphi q}^{(3)}-c_{\varphi d}\big) \, .
\end{align}
Putting all together allows one to compute the SMEFT effects to the $Z$ pole observables measured at LEP, which results in the constraint of $7$ directions in the parameter space spanned by the operators in Tab.~\ref{tab:ops}.

Another important observable to consider is Bhabha scattering, the process $e^+ e^- \to e^+ e^-$, which played a crucial role in calibrating luminosity at LEP while also providing stringent constraints on new physics. The high-statistics and precision of these measurements enabled detailed studies of the electroweak couplings of electrons. Within the SMEFT framework, flavour-blind operators can induce modifications to these couplings, leading to measurable deviations in angular distributions and total cross sections in Bhabha scattering. Incorporating this observable extends the sensitivity to an additional direction in parameter space, leaving only the TGCs as unprobed. The same effect can be obtained with the inclusion of the measurement of $\alpha_{EW}$.

Finally, the WW diboson production, $e^+ e^- \to W^+ W^-$, constitutes another significant dataset from LEP, crucial for testing the non-abelian structure of the SM's electroweak sector. Precision measurements of the production cross section and angular distributions provide stringent tests of the SM predictions. In the SMEFT framework, deviations in these observables can stem from operators that modify the couplings to the $Z$ boson, alter gauge-boson self-interactions in the SM, or introduce higher-dimensional Lorentz structures.
Specifically, the modified TGC Lagrangian reads
\begin{equation}\label{eq:WWW}
    \frac{\mathcal{L}_{W W V}}{-i g_{W W V}}=g_1^V\left(W_{\mu \nu}^{+} W^{-\mu} V^\nu-W_\mu^{+} V_\nu W^{-\mu \nu}\right)+\kappa_V W_\mu^{+} W_\nu^{-} V^{\mu \nu}+\frac{i \delta\lambda_V}{m_W^2} V^{\mu \nu} W_\nu^{+\rho} W_{\rho \mu}^{-} \, ,
\end{equation}
where $V = Z, \gamma$ and we have defined $V_{\mu \nu}=\partial_\mu V_\nu-\partial_\nu V_\mu$ and $W_{\mu \nu}^{ \pm}=\partial_\mu W_\nu^{ \pm}-\partial_\nu W_\mu^{ \pm}$.

The SM couplings are given by
\begin{align}
    g_{WW\gamma} = e \, , \qquad
    g_{WWZ} = e \cot{\theta_W} \, , 
\end{align}
where the electric charge $e$ and the trigonometric functions of the Weinberg angle can be determined in terms of the chosen EW input parameters.

Five of the dimension-six SMEFT operators in Table~\ref{tab:ops} introduce a dependence on the couplings $g_1^V = 1 + \delta g_1^V$ and $\kappa_V = 1 + \delta \kappa_V$, i.e.
\begin{align}\begin{aligned}\label{eq:deltag}
\delta g_1^\gamma & =\frac{1}{4 \sqrt{2} G_F}\left(c_{\varphi D} \frac{m_W^2}{m_W^2-m_Z^2}-4 c_{\varphi l}^{(3)}+2 c_{ll}-c_{\varphi W B} \frac{4 m_W}{\sqrt{m_Z^2-m_W^2}}\right) \, , \\
\delta g_1^Z & =\frac{1}{4 \sqrt{2} G_F}\left(c_{\varphi D}-4 c_{\varphi l}^{(3)}+2 c_{ll}+4 \frac{m_Z}{m_W} \sqrt{1-\frac{m_W^2}{m_Z^2}} c_{\varphi W B}\right) \, , \\
\delta \kappa_\gamma & =\frac{1}{4 \sqrt{2} G_F}\left(c_{\varphi D} \frac{m_W^2}{m_W^2-m_Z^2}-4 c_{\varphi l}^{(3)}+2 c_{ll}\right) \, , \\
\delta \kappa_Z & =\frac{1}{4 \sqrt{2} G_F}\left(c_{\varphi D}-4 c_{\varphi l}^{(3)}+2 c_{ll}\right) \, .
\end{aligned}\end{align}
It is important to highlight that the coefficients $c_{\varphi l}^{(3)}$, $c_{ll}$, and $c_{\varphi D}$ induce universal modifications to the TGCs within the SM. In contrast, the coefficient $c_{\varphi W B}$ does not influence $\kappa_V$ but exclusively alters $g_1^V$, thereby breaking the symmetric structure characteristic of SM interactions.

Of the operators under consideration, only one gives rise to a novel Lorentz structure by generating a contribution proportional to $\delta \lambda_V$. This dependence is expressed as:
\begin{align}
    \delta \lambda_\gamma &= -6 \sin{\theta_W} \frac{m_W^2}{g_{WW\gamma}} c_W \, , \\
    \delta \lambda_Z &= -6 \cos{\theta_W} \frac{m_W^2}{g_{WWZ}} c_W \, .
\end{align}
The inclusion of $WW$ production at LEP allows to probe these remaining parameter space dimensions and constrain the full set of operators considered in this study.

\begin{table}[t!]
    \begin{center}
  {\fontsize{8pt}{8pt}\selectfont
    \centering
     \renewcommand{\arraystretch}{2}
     \setlength{\tabcolsep}{10pt}
     \begin{tabular}{lcccc}
       \toprule \textbf{Exp.}   & $\bf{\sqrt{s}}$ \textbf{(GeV)}
      &  \textbf{Observable} & $\mathbf{n_{\rm dat}}$
       &\textbf{Ref.}\\
      \toprule
        \bf{LEP}
        & 91
        & Z observables
        & 19
        & \cite{ALEPH:2005ab}\\
      \midrule
        \bf{LEP}
        & 80.386
        & $\mathcal{B}(W \rightarrow l^{-} \bar{v}_l)$
        & 3
        & \cite{ALEPH:2013dgf}\\
      \midrule
        \bf{LEP}
        & 189-207
        & $\sigma(e^+ e^- \rightarrow e^+ e^-)$
        & 21
        & \cite{ALEPH:2013dgf}\\
      \midrule
        \bf{LEP}
        & 91
        & $\hat{\alpha}^{(5)}_{\rm}(M_Z)$
        & 1
        & \cite{Workman:2022ynf}\\
         \midrule
        \bf{LEP}
        & 182 - 206
        & $d \sigma _{WW} / d cos(\theta _W)$
        & 40
        & \cite{ALEPH:2013dgf}\\
  \bottomrule
     \end{tabular}
     \vspace{0.3cm}
     \caption{Measurements of electroweak precision observables included in the analysis. The columns contain information on the 
     centre-of-mass energy, the observable, the number 
     of data points and the source.
      \label{tab:dataset_ewpo}
  }
  }
  \end{center}
  \end{table}

\begin{figure}[t!]
    \centering
    \includegraphics[width=0.95\linewidth]{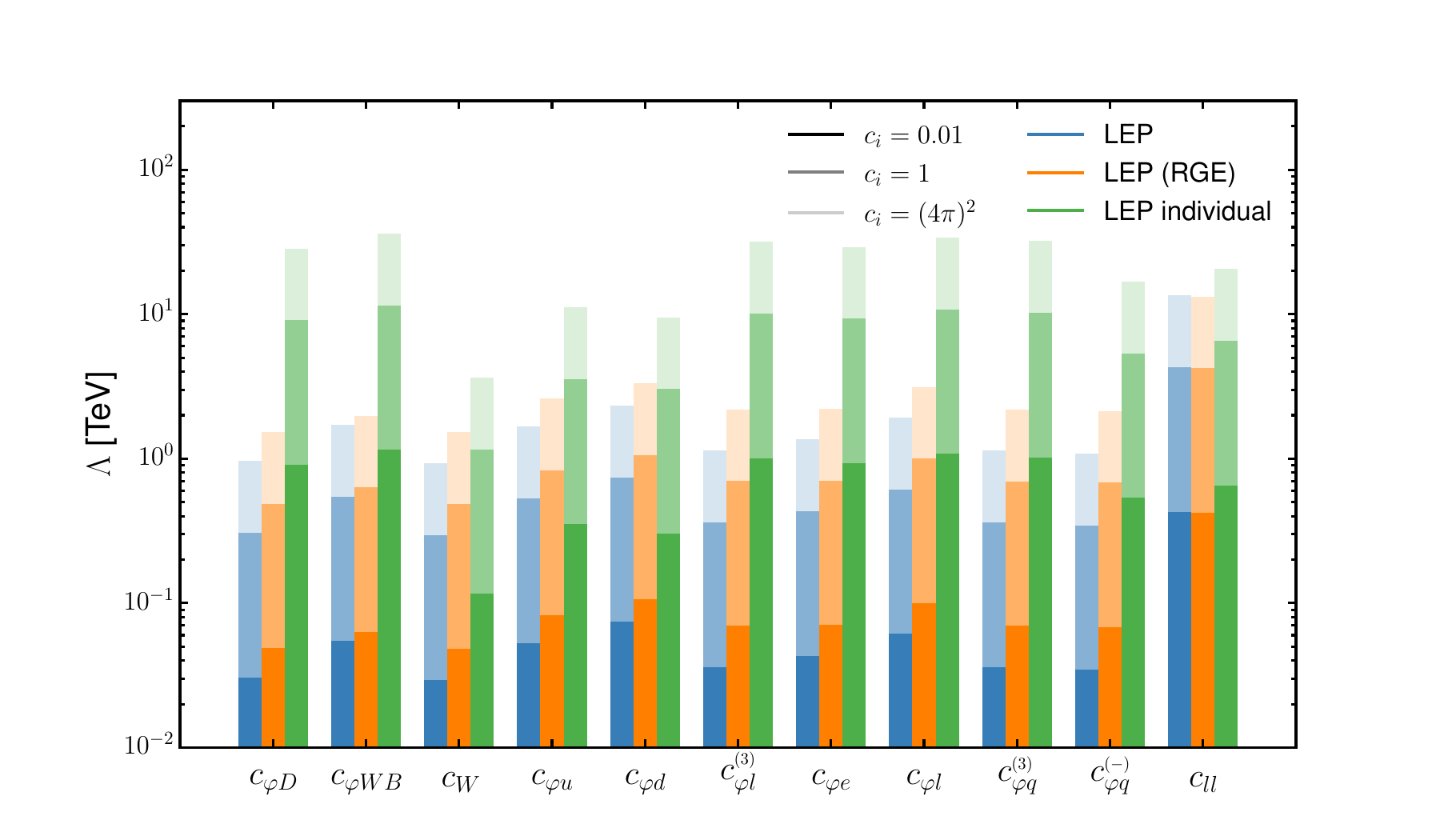}
    \caption{Bar plots showing the mass reach $\Lambda$ for each Wilson coefficient in Table~\ref{tab:ops} using LEP data. Three different scenarios are presented: in blue, the marginalized constraints without RGE effects; in orange, the constraints obtained when including RGE effects from a reference scale of $\mu_0 = 1$ TeV; and in green, the results from individual fits.}
    \label{fig:LEP-fits}
\end{figure}

We collect in Tab.~\ref{tab:dataset_ewpo} all the data from LEP measurements that we employ in the present study.
LEP data imposes very stringent constraints on UV models. However, when adopting an agnostic approach that explores effects from all possible Wilson coefficients, the constraining power becomes significantly diluted, particularly for certain operators. This effect is clearly illustrated in Fig.~\ref{fig:LEP-fits}, where we compare the $95\%$ C.L. marginalised bounds on various operators constrained by LEP data (with and without RGE effects) with those obtained from individual fits. 

The bounds are expressed in terms of the mass reach at $95\%$ C.L., where we define:
\begin{equation}
\Lambda = \sqrt{\frac{C_i}{2 \delta_{c_i}}}
\label{eq:lambda_def}
\end{equation}
with $\delta_{c_i}$ denoting the uncertainty on the ratio $C_i/\Lambda^2$ as determined by the data, i.e. the marginal standard deviation of each Wilson coefficient. We present $\Lambda$ values for three benchmark scenarios: $C_i = 0.01, 1, (4\pi)^2$, corresponding to weakly coupled, intermediately coupled, and strongly coupled NP, respectively. Note that this way of presenting results is compressing a bit the differences due to the square root in Eq.~\eqref{eq:lambda_def}, e.g. if the bounds on the Wilson coefficients improve by a factor four, the resulting mass reaches will only improve by a factor two.

While marginal bounds are generally expected to be weaker than individual bounds, the stark difference observed in the case of LEP indicates that the considered observables cannot fully disentangle different parameter space directions. This suggests that strong correlations between operators significantly hinder the constraining power. 

\begin{figure}[t!]
    \centering
    \includegraphics[width=0.45\linewidth, page=8]{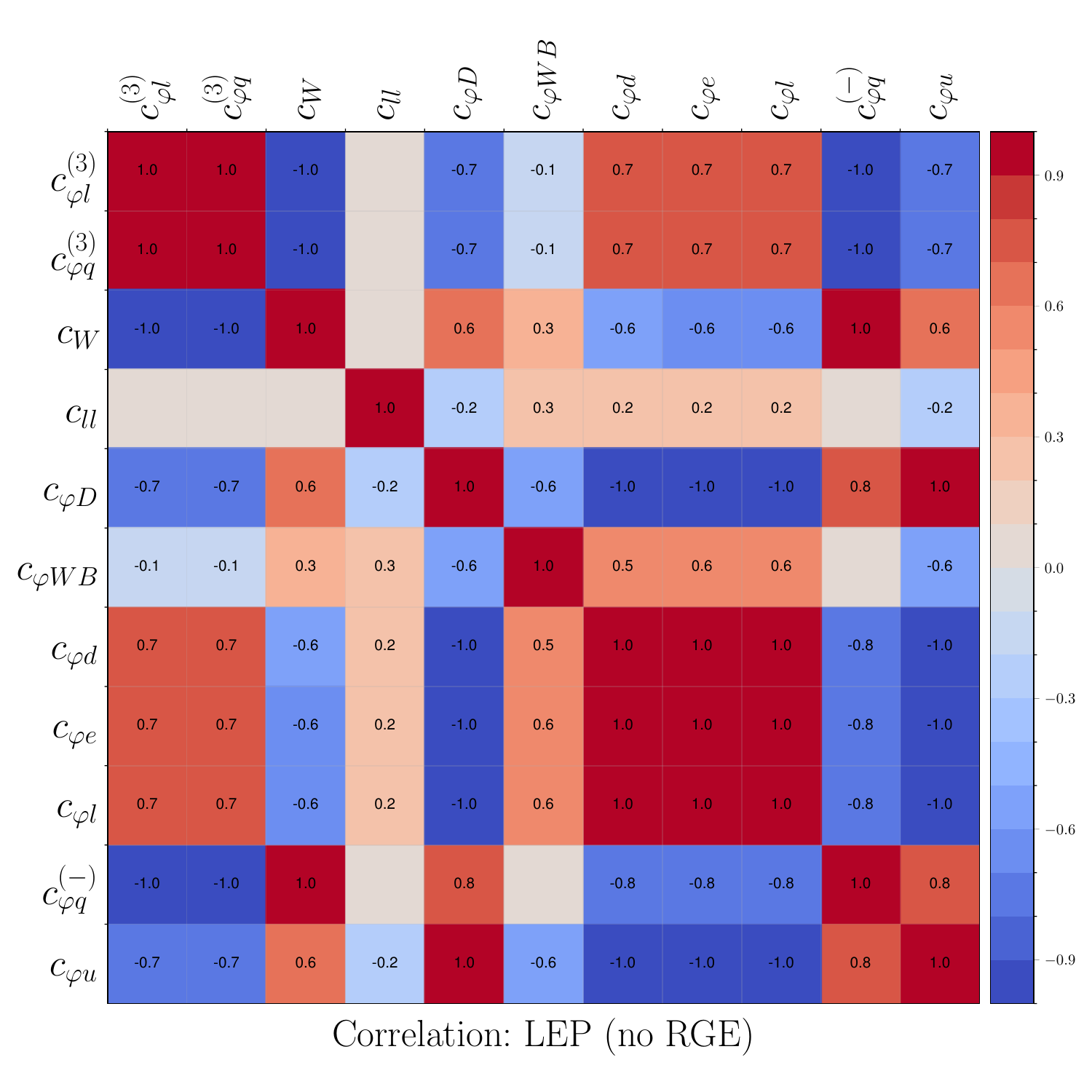}
    \includegraphics[width=0.45\linewidth, page=9]{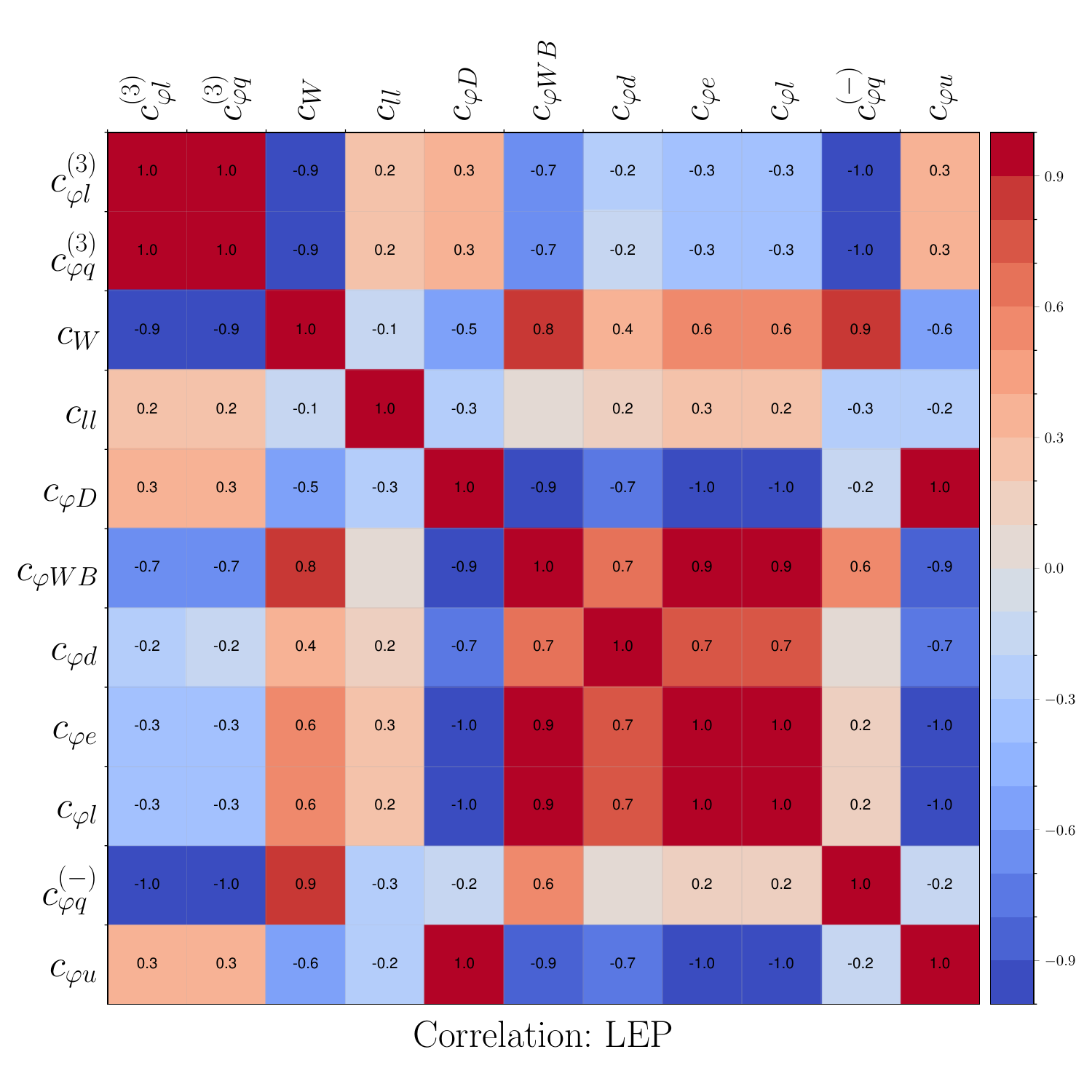}
  \caption{Correlation matrices among the 11 relevant SMEFT operators for the LEP only fits. The left panel shows results using legacy LEP data while the right panel shows the correlations for a fit including RGE effects. This highlights how the inclusion of scale dependence in the Wilson coefficients affect correlations between operators, reducing degeneracies and improving the robustness of the fit.}
\label{fig:correlation_matrices_LEP}
\end{figure}

This issue is made explicit in Fig.~\ref{fig:correlation_matrices_LEP}, where the left panel displays the correlation matrix for a global fit involving eleven operators constrained by LEP data. The presence of high correlation values signals quasi-degeneracies in the parameter space, highlighting that additional observables have the potential to introduce important new sensitivities. By effectively unlocking the potential of LEP data, these new inputs could lead to significant improvements in the fit beyond naive expectations.

Interestingly, once RGE effects are incorporated into the predictions, several of the previously strong correlations are at least partially reduced, as shown in the right panel of Fig.~\ref{fig:correlation_matrices_LEP}. Consequently, the bounds on the Wilson coefficients exhibit a noticeable improvement (see Fig.~\ref{fig:LEP-fits}, orange bars). However, further refinements are possible, as a significant gap remains compared to the individual bounds. Notably, RGE effects can also enhance correlations for certain Wilson coefficients, such as the pair $c_{\varphi W B} - c_{W}$, a common outcome of their inclusion that often weakens constraints. Nevertheless, this is not the case for the LEP fit considered here, which, after incorporating RGE effects, shows a general improvement in constraints.
In particular, this effect is independent of the reference scale $\mu_0$ chosen for the Wilson coefficients. In the plots, we set $\mu_0 = 1$ TeV, but a similar qualitative result is obtained if instead $\mu_0 = m_Z$ is chosen.
\begin{figure}[t!]
    \centering
    \includegraphics[width=0.9\linewidth]{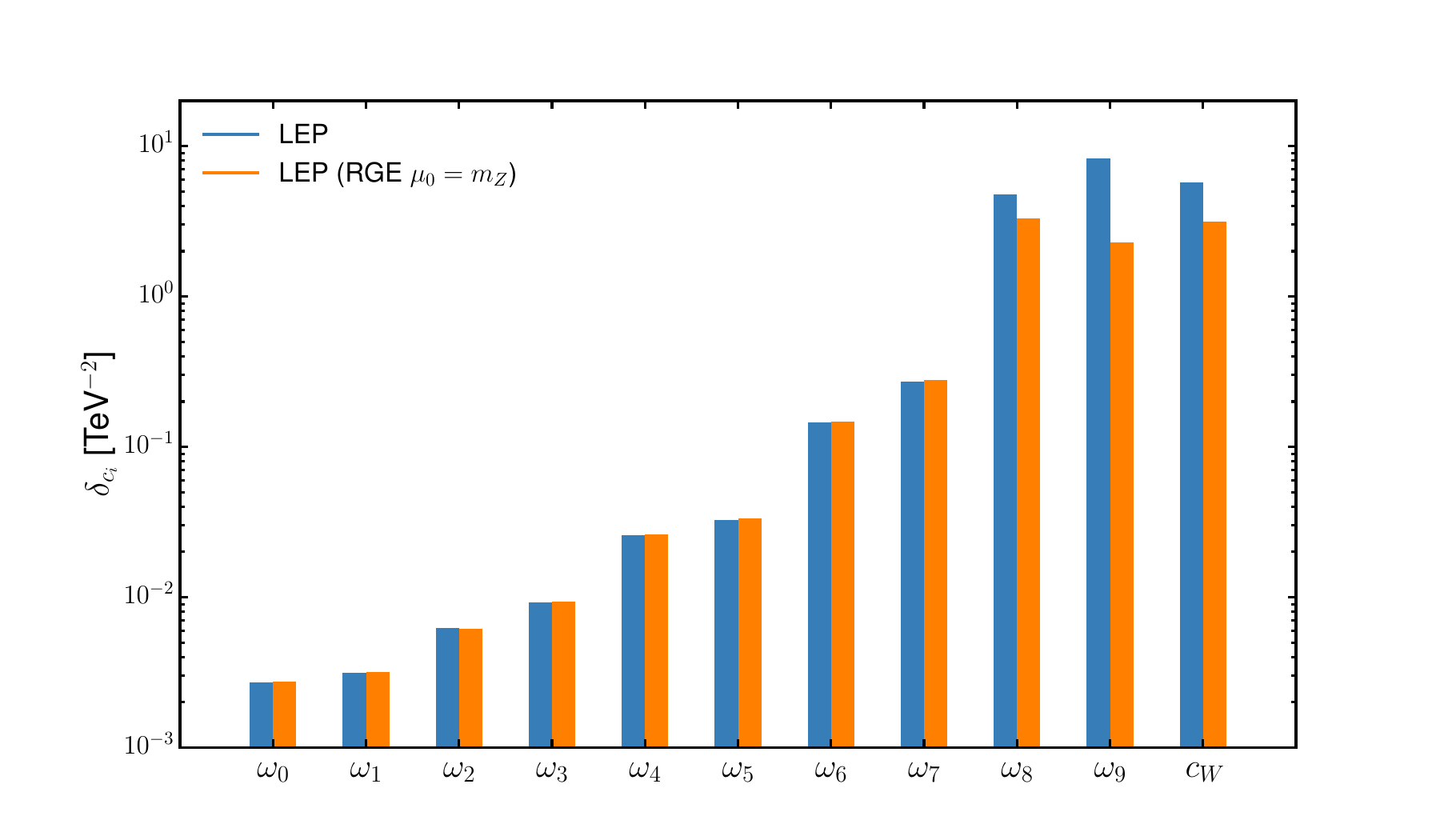}
    \caption{Bar plot showing the uncertainties on the principal components of the $Z$-pole observables obtained from a global LEP fit. The results without RGE effects are shown in blue, while those including RGE effects with a reference scale of $\mu_0 = m_Z$ are shown in orange.}
    \label{fig:unc_plot_pca}
\end{figure}

\begin{figure}[t!]
    \centering
    \includegraphics[width=0.7\linewidth]{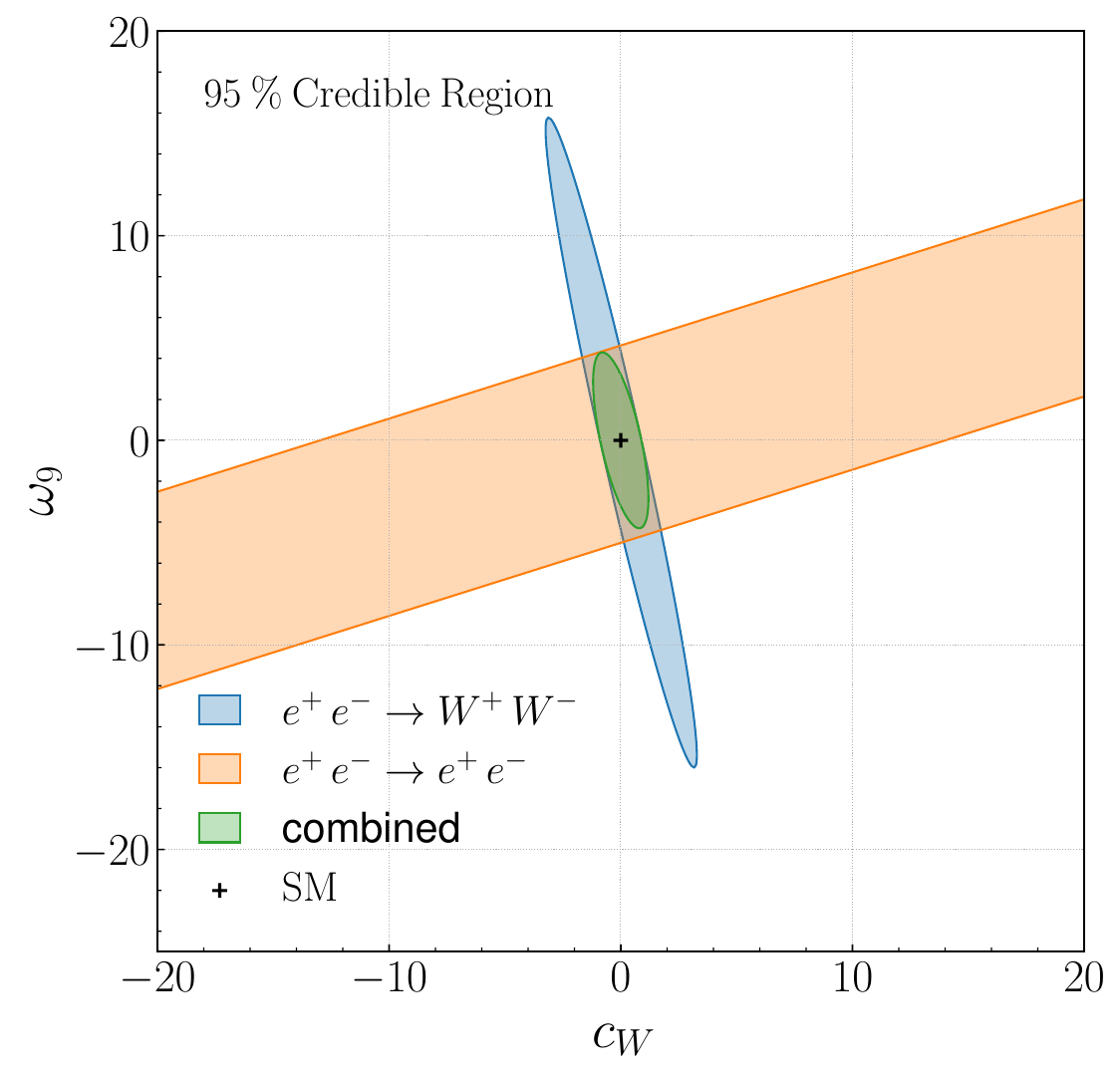}
    \caption{Credible regions at $95\%$ C.L. for a two-dimensional fit of the principal component $\omega_9$ and the triple gauge coupling Wilson coefficient $c_W$. The bounds obtained from diboson data are shown in blue, those from Bhabha scattering data in orange, and their combination in green. All constraints include RGE effects, with the Wilson coefficients defined at the reference scale $\mu_0 = m_Z$.}
    \label{fig:w9_cW}
\end{figure}
In order to understand why the RGE improves the bounds, it is useful to perform a change of basis. In particular, considering the Wilson coefficients defined at the reference scale $\mu_0 = m_Z$, a natural parametrization of the EFT effects is given by the Principal Component Analysis (PCA) of the Z-pole data and the measurement of $\alpha_{EW}$. The definition of the PCA directions is reported in App.~\ref{app:pca}.
This dataset is sensitive to 10 operators in the flavour-blind scenario, namely all the operators listed in Table~\ref{tab:ops}, except for $\mathcal{O}_{WWW}$. However, it is well known that only 8 independent directions in this parameter space can be constrained~\cite{Brivio:2017bnu}. Performing a PCA reveals the orthogonal directions dictated by these data, denoted here as $\omega_i$ with $i=0,\dots,9$. In particular, the last two directions, $\omega_8$ and $\omega_9$, are in direct relation with the TGCs. This explains, from a physical perspective, why they cannot be constrained by Z-pole observables alone, necessitating the inclusion of diboson production data to bound them effectively. 

In Fig.~\ref{fig:unc_plot_pca}, we present the uncertainties $\delta_{c_i}$ on the various principal components obtained from a fit to the full LEP dataset, both without RGE effects and with their inclusion. Notably, in this setup, the reference scale is $\mu_0 = m_Z$, implying a moderate scale separation between the Z-pole and the energy scales probed by diboson production and Bhabha scattering, which are on the order of 200 GeV. Despite this relatively small separation, we observe that the RGE leads to a significant improvement in the bounds on $\omega_8$, $\omega_9$, and $c_W$, the directions associated with the TGCs. 

This improvement arises because, in the presence of running effects, Bhabha scattering becomes sensitive to the TGCs through loop effects and helps break a strong correlation in the $\omega_9 - c_W$ plane, which is present when only diboson data is considered (see Fig.~\ref{fig:w9_cW}, where the fit is done assuming that the data lies exactly on the SM for illustration purposes). This effect is particularly striking, as Bhabha scattering has minimal impact on the fit in the absence of RGE effects, leading one to believe it could be safely neglected due to its limited information content. However, once RGE effects are taken into account, Bhabha scattering becomes crucial, enabling a significant enhancement in the constraints of anomalous TGCs. 

This observation further underscores the importance of global SMEFT interpretations that incorporate as much experimental data as possible. It also highlights how advancements in theoretical predictions can yield unexpected improvements in sensitivity, unlocking new avenues for constraining new physics beyond naive expectations.

%% file: sec-results.tex
\section{Beyond LEP: synergies with complementary data}
\label{sec:globalfits}
In this section, we examine the interplay between LEP data and other observables that probe complementary directions in parameter space. As discussed in the previous section, LEP data fits exhibit strong correlations, suggesting that significant improvements in constraining the operators should be possible. To enhance the analysis, we incorporate data from two different energy regimes, allowing us to explore their distinct implications.

First, we examine the impact of diboson production at the LHC, which serves as a crucial probe of TGC at energies significantly higher than those accessed by LEP. Incorporating these datasets is essential for breaking correlations in traditional SMEFT analyses at the LHC. Second, we explore the insights gained from low-energy observables in the flavour sector. The diverse sources of constraints, particularly the energy scales they probe, have profound implications for the validity of SMEFT analyses.

\subsection{Diboson production at the LHC}
\label{sec:diboson}
The typical complementary probe of TGCs considered in the literature is diboson production at the LHC. In particular, it has been shown that the production of $W^+ W^-$ and $WZ$ is significantly more sensitive to anomalous couplings than the data from LEP~\cite{Butter_2016}. However, it is important to stress that this statement holds only when considering the strong constraints on gauge couplings to fermions derived from LEP data, while neglecting their correlations with TGCs in the analysis. In fact, if one performs an agnostic fit to diboson data, allowing all couplings to float independently, the sensitivity to TGCs is actually worse than that obtained from $W^+ W^-$ production at LEP.

\begin{table}[t!]
    \begin{center}
  {\fontsize{8pt}{8pt}\selectfont
    \centering
     \renewcommand{\arraystretch}{2}
     \setlength{\tabcolsep}{10pt}
     \begin{tabular}{lcccc}
       \toprule \textbf{Exp.}   & $\bf{\sqrt{s}}$ \textbf{(TeV)}
      &  \textbf{Observable} & $\mathbf{n_{\rm dat}}$
       &\textbf{Ref.}\\
      \toprule
        \bf{ATLAS}
        & 13
        & $d \sigma _{W^+W^-}/d m_{e \mu}$
        & 13
        & \cite{ATLAS:2019rob}\\
      \midrule
        \bf{ATLAS}
        & 13
        & $d \sigma _{WZ} / d m_{T}$
        & 6
        & \cite{ATLAS:2019bsc}\\
      \midrule
        \bf{CMS}
        & 13
        & $d \sigma _{WZ} / d p^Z_{T}$
        & 11
        & \cite{CMS:2019efc}\\
       \midrule
        \bf{CMS}
        & 13
        & $1/\sigma \, d \sigma _{WZ} / d p^Z_{T}$
        & 11
        & \cite{CMS:2021icx}\\
  \bottomrule
     \end{tabular}
     \vspace{0.3cm}
     \caption{Measurements of diboson production at the LHC included in the analysis. The columns contain information on the 
     centre-of-mass energy, the observable, the number 
     of data points and the source.
      \label{tab:dataset_diboson}
  }
  }
  \end{center}
  \end{table}
\begin{figure}[t!]
    \centering
    \includegraphics[width=0.9\linewidth]{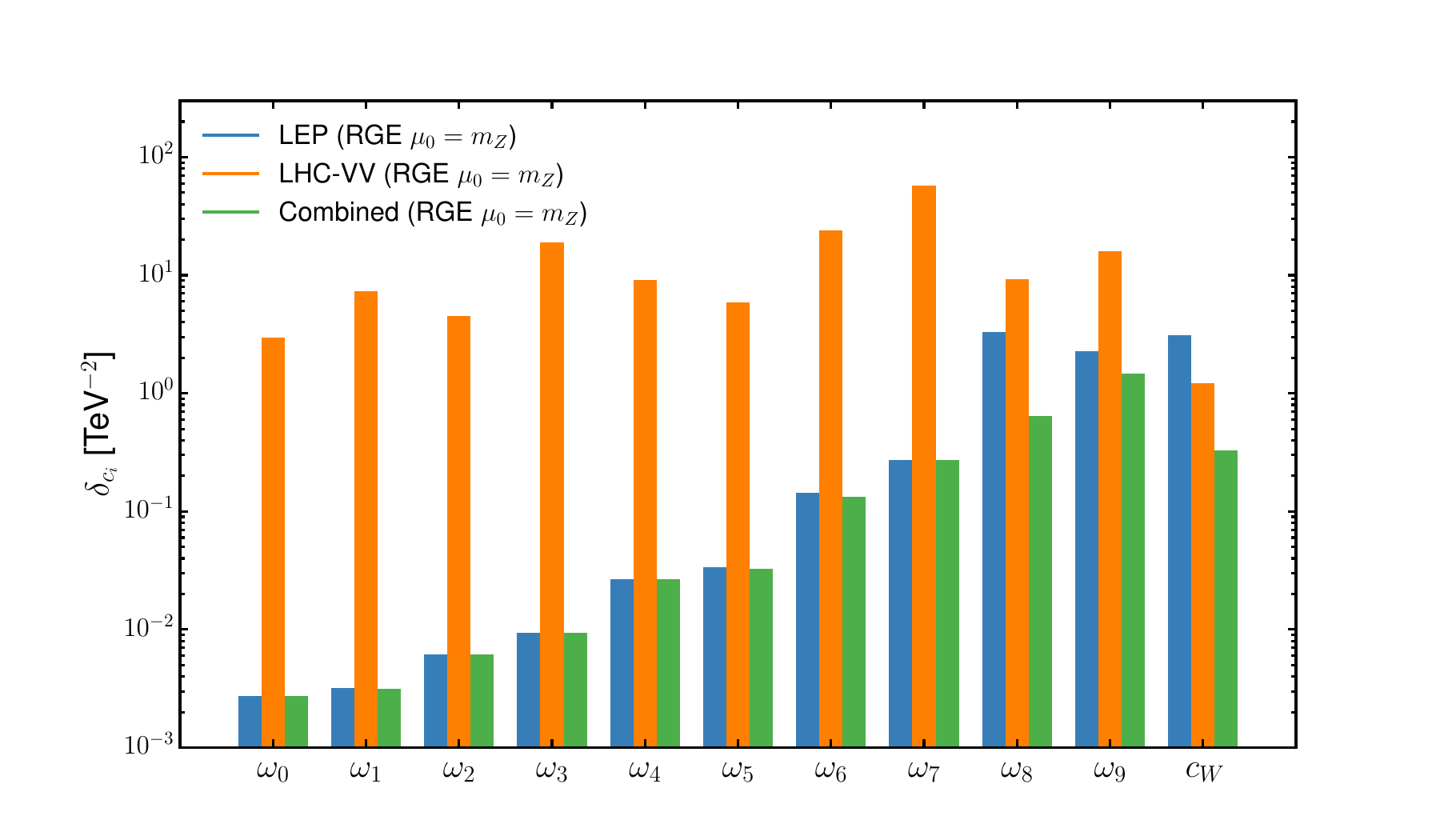}
    \caption{Bar plot showing the marginal uncertainties on the principal components of the Z-pole
observables obtained by fitting LEP and diboson data. The results for a LEP only fit are shown
in blue, the ones for a LHC diboson fit only in orange and their combination in green. RGE effects are accounted for and the Wilson coefficients are defined at the reference scale $\mu_0 = m_Z$.}
    \label{fig:LEP_vs_LHC-VV}
\end{figure}
\begin{figure}[t!]
    \centering
    \includegraphics[width=0.7\linewidth]{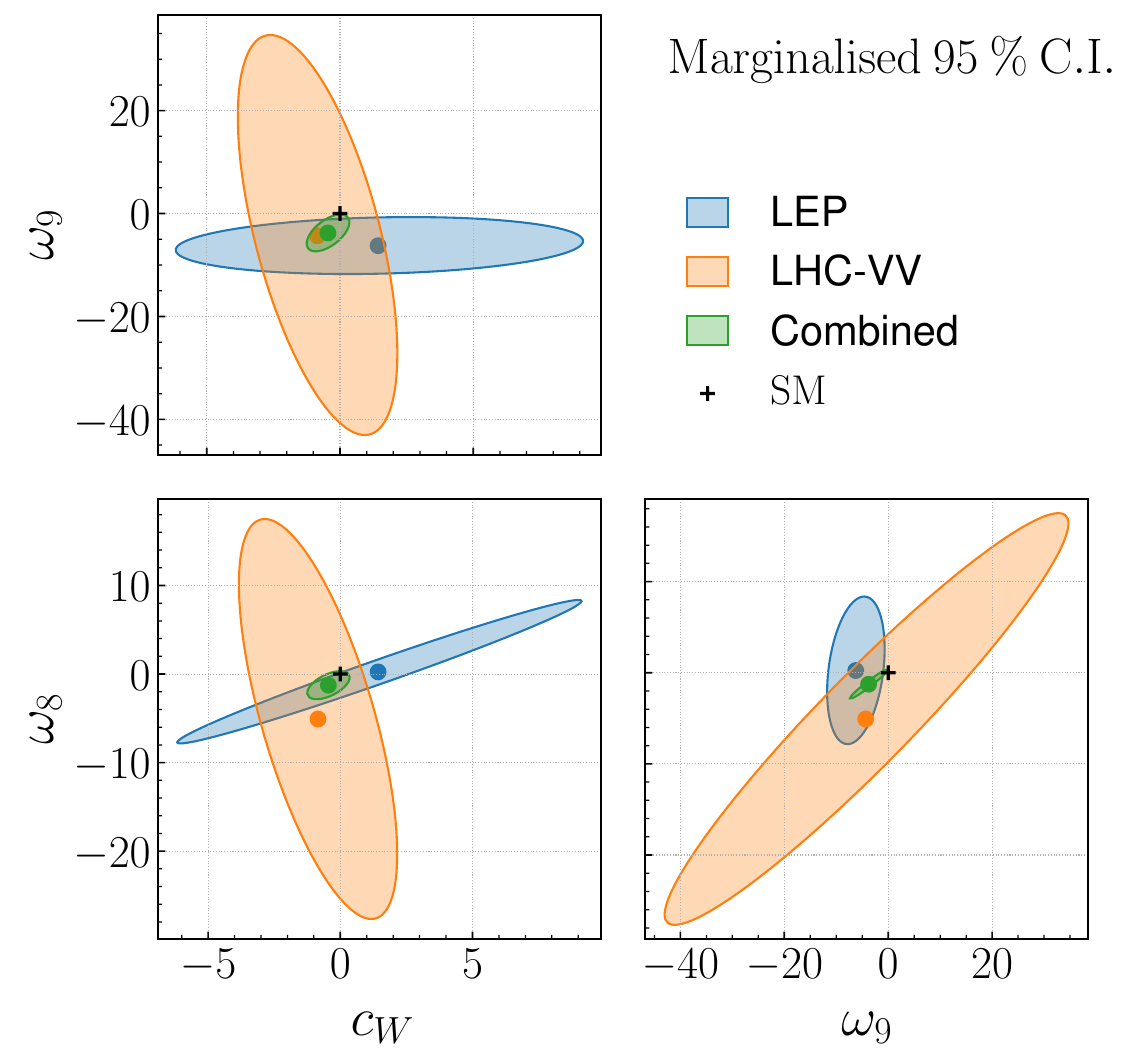}
    \caption{Marginalised credible intervals of the $\omega_8$, $\omega_9$ and $c_W$ Wilson coefficients obtained from LEP and and LHC diboson fits. In all cases RGE effects are accounted for and the Wilson coefficients are defined at the reference scale $\mu_0=m_Z$.}
    \label{fig:LEP_vs_LHC-VV_2D}
\end{figure}
In the following, we will consider diboson production at the LHC as measured by ATLAS and CMS, relying on the dataset from the \smefit analysis in Ref.~\cite{Celada:2024mcf}. The differential distributions provided therein probe energies ranging from $\sim 100$ GeV to the TeV scale. The considered datasets are provided in Table~\ref{tab:dataset_diboson}. We compared our results with the recent analysis in Ref.~\cite{Celada:2024cxw}, which also assesses the impact of triboson production on SMEFT global fits. We find good agreement in the fits at LEP while some quantitative differences are observed when diboson data is added. The main differences are arising from the inclusion of the $\Delta \phi_{jj}$ differential distribution in $Zjj$ production in the latter analysis, which significantly enhances the constraining power on $c_W$. Additionally, our analysis incorporates RGE effects, which have a relatively minor impact on the overall constraining power but lead to a more pronounced shift in the best-fit point.

In Fig.~\ref{fig:LEP_vs_LHC-VV}, we present the marginal uncertainties obtained by fitting the PCA directions determined at the $Z$-pole, as defined in App.~\ref{app:pca}. Recall that the directions $\omega_0$ to $\omega_7$ correspond to the couplings of fermions to gauge bosons, while $\omega_8$ and $\omega_9$ are associated with the TGCs. The figure illustrates that only by combining the information from LEP and diboson data does the sensitivity to TGCs improve significantly. The only Wilson coefficient that already benefits from an enhancement at the level of diboson datasets alone is $c_W$. 

It is worth noting that this conclusion is further reinforced by the inclusion of RGE effects in the fit. As discussed in the previous section, these effects lead to a substantial improvement in the TGC constraints at LEP, primarily due to the additional sensitivity introduced by Bhabha scattering data.

To explicitly illustrate how diboson production leads to a significant improvement in the bounds on TGCs, Fig.~\ref{fig:LEP_vs_LHC-VV_2D} presents a corner plot of the marginalized two-dimensional credible regions for the $\omega_8$, $\omega_9$, and $c_W$ coefficients. Note that these regions do not result from a fit performed on only these three coefficients; rather, they correspond to projections onto the two-dimensional parameter space from the full global fit. 

While some of the cause for improvement is obscured by the marginalization procedure, the plot clearly demonstrates the complementarity between the two datasets. That is, LEP and diboson data probe different directions in parameter space, leading to a substantial enhancement in the overall constraint power when combined.

Despite the powerful bounds that can be obtained from the interplay of LEP and LHC diboson data, concerns are often raised when relatively high-energy data is used within an EFT approach. While it is true that high energies typically lead to a substantial increase in sensitivity due to the energy enhancements provided by higher-dimensional operator insertions in amplitudes~\cite{Farina:2016rws}, they also weaken the model independence of the analysis. In particular, reinterpretations in terms of UV models can be severely limited, as the validity of the EFT framework may be violated. This occurs, for instance, when the masses of new particles are close to the energy scales probed by the data, in which case a reliable low-energy description would require the inclusion of higher-dimensional operators, such as those of dimension $8$ or beyond~\cite{Dawson:2023ebe}.

These concerns are often addressed by extending the EFT expansion, either by incorporating partial $\mathcal{O}(1/\Lambda^4)$ corrections through the inclusion of squared dimension-six amplitudes or by performing full dimension-$8$ calculations. While the latter approach extends the validity of the EFT framework, it comes at the cost of significantly increasing the number of parameters under consideration, which is generally regarded as prohibitive in an agnostic global fit.

An alternative strategy is to restrict the analysis to data that does not probe energies beyond a certain threshold, ensuring that the resulting bounds remain robust for reinterpretation in UV models. In the following section, we explore this possibility in greater detail, replacing the LHC diboson data with flavour observables instead. This approach ensures a much broader reliability of the obtained bounds while maintaining EFT validity.

\subsection{The impact of flavour observables}
\label{sec:flavour_observables}
\begin{figure}[h!]
    \centering
    \includegraphics[width=0.35\linewidth]{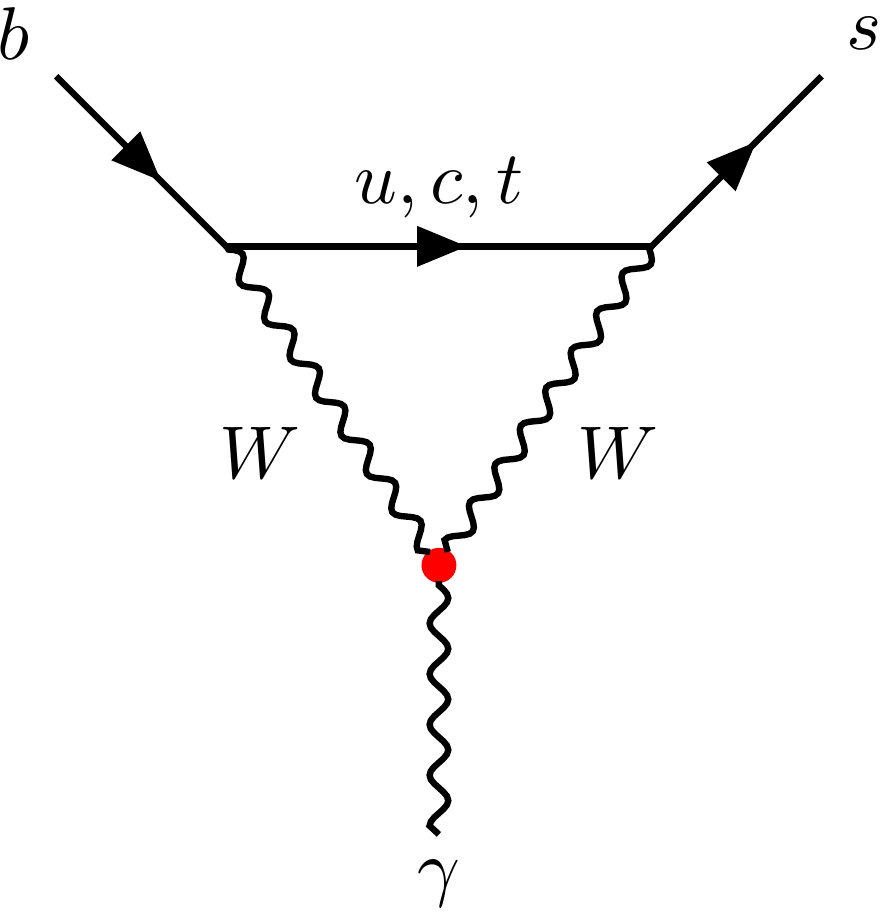} \; \;
    \includegraphics[width=0.35\linewidth]{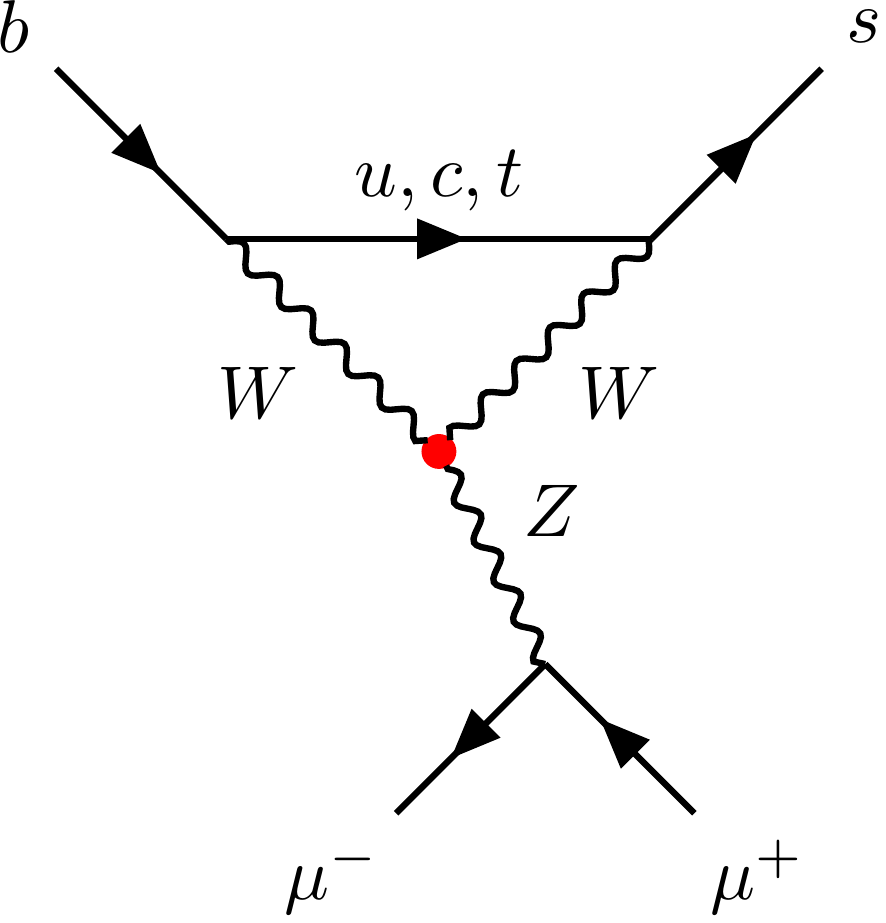}
    \caption{Feynman diagrams illustrating the loop-induced transitions $b \to s \gamma$ and $b \to s \mu^+ \mu^-$, emphasizing their dependence on the TGC, marked with a red dot. Anomalous interactions modify these vertices, leading to potential deviations in the corresponding observables.}
    \label{fig:flav_tgc}
\end{figure}

Flavour observables provide a powerful testing ground for the SMEFT, as they are sensitive to new physics effects at both low and high energies. Flavour-blind SMEFT operators, which assume no explicit breaking of the SM's flavour symmetries, can subtly influence flavour observables in measurable ways. These effects often manifest in rare decays, mixing phenomena, and CP violation, where precision measurements impose stringent constraints on new physics scenarios.

Recent studies have investigated the interplay between flavour-blind SMEFT contributions and flavour observables. Notably, the analyses in Refs.~\cite{Greljo:2023adz, Greljo:2023bdy} explore how flavour-blind scenarios can be constrained using flavour observables, assessing also the importance of RGE effects. Additionally, in Refs~\cite{Greljo:2023bab, Greljo:2022cah, Greljo:2022jac}, the authors examine the impact of flavour observables and their connection to high-energy data, offering valuable insights into how SMEFT effects propagate across different energy scales.

Global fits have also played a crucial role in investigating the relationship between flavour and high-energy observables~\cite{Bartocci:2023nvp, Bruggisser:2022rhb, Aoude:2020dwv, Bruggisser:2021duo}. These studies consider various flavour scenarios, with a particular emphasis on the framework of Minimal Flavour Violation (MFV). By combining flavour and collider data, such global fits enhance our ability to constrain SMEFT parameter spaces, underscoring the complementarity of low- and high-energy measurements.

Flavour observables are conventionally described within the WET framework. In this study, we do not aim for an exhaustive analysis but instead select a subset of observables that allow us to probe all WET coefficients discussed in Sec.~\ref{sec:theory}, namely $C_7$, $C_9$, and $C_{10}$. The chosen observables are:

\begin{itemize}
    \item \bm{$B \to X_s \gamma$}: This process corresponds to the inclusive radiative decay of a $B$ meson into a strange quark and a photon. As a FCNC transition occurring at the loop level in the SM, it is highly sensitive to new physics contributions. Precise measurements of its branching ratio impose stringent constraints on scenarios involving heavy new particles that couple to the top quark and electroweak gauge bosons. This transition primarily probes the Wilson coefficient $C_7$.

    \item \bm{$B_s \to \mu^+ \mu^-$}: The rare decay of the $B_s$ meson into a muon-antimuon pair is a FCNC process that is strongly suppressed in the SM. It proceeds via loop and box diagrams, making it highly sensitive to new physics contributions, particularly from scalar or pseudoscalar interactions. The SM predicts its branching ratio with high precision, and deviations from experimental measurements could indicate physics beyond the SM, such as additional Higgs bosons, leptoquarks, or heavy gauge bosons. This process is predominantly sensitive to $C_{10}$.

    \item \bm{$P_5^\prime$}: This angular observable is extracted from the differential decay distribution of $B^0 \to K^{*0} \mu^+ \mu^-$. Designed to minimize hadronic uncertainties, it provides a theoretically clean probe of potential new physics contributions~\cite{Matias:2012xw}. Deviations from SM predictions in $P_5^\prime$ have been reported in experimental data~\cite{LHCb:2020lmf}, generating significant interest in possible explanations involving modified Wilson coefficients or new heavy mediators. This observable is sensitive to all three coefficients, $C_7$, $C_9$, and $C_{10}$, making it particularly valuable for testing new physics scenarios.
\end{itemize}

Overall, $B \to X_s \gamma$ provides a direct probe of $C_7$, $B_s \to \mu^+ \mu^-$ is predominantly sensitive to $C_{10}$, and $P_5^\prime$ serves as a key probe of $C_9$ while also being affected by $C_7$ and $C_{10}$. Together, these observables offer a comprehensive test of flavour physics within the flavour-blind SMEFT framework in our study. 
We incorporate these observables by interfacing with \flavio~\cite{Straub:2018kue}, utilising its likelihood as a function of $C_7$, $C_9$, and $C_{10}$ as an external likelihood in \smefit, where a custom map from the \smefit Wilson coefficient basis to the WET coefficients is implemented using the matching expressions provided in Ref.~\cite{Hurth:2019ula}.
We have verified independently that the combination of these 3 observables gives qualitatively good constraints on the WET coefficients, resulting in comparable bounds to the ones obtained in a similar study~\cite{Aoude:2020dwv}.

In particular, these observables are highly sensitive to the TGCs, as illustrated in Fig.~\ref{fig:flav_tgc}. This was the focus of Ref.~\cite{Bobeth:2015zqa}, which demonstrated that, under certain assumptions, radiative and rare B-meson decays impose constraints on two of the three anomalous couplings in Eq.~\eqref{eq:deltag}, achieving a level of precision comparable to that obtained from LEP II, Tevatron, and LHC data. 

The goal of the present analysis is to extend this study by incorporating updated measurements and demonstrating that the inclusion of flavour observables within a global fit framework can further enhance sensitivity. Specifically, we show that flavour data can break parameter correlations and push the mass reach beyond LEP precision limits. Unlike diboson analyses, this improvement is achieved without relying on the high-energy enhancement of the LHC, which could otherwise challenge the validity of the EFT expansion. This ensures a more robust reinterpretation of the results in terms of UV completions.

\begin{figure}[t!]
    \centering
    \includegraphics[width=0.9\linewidth]{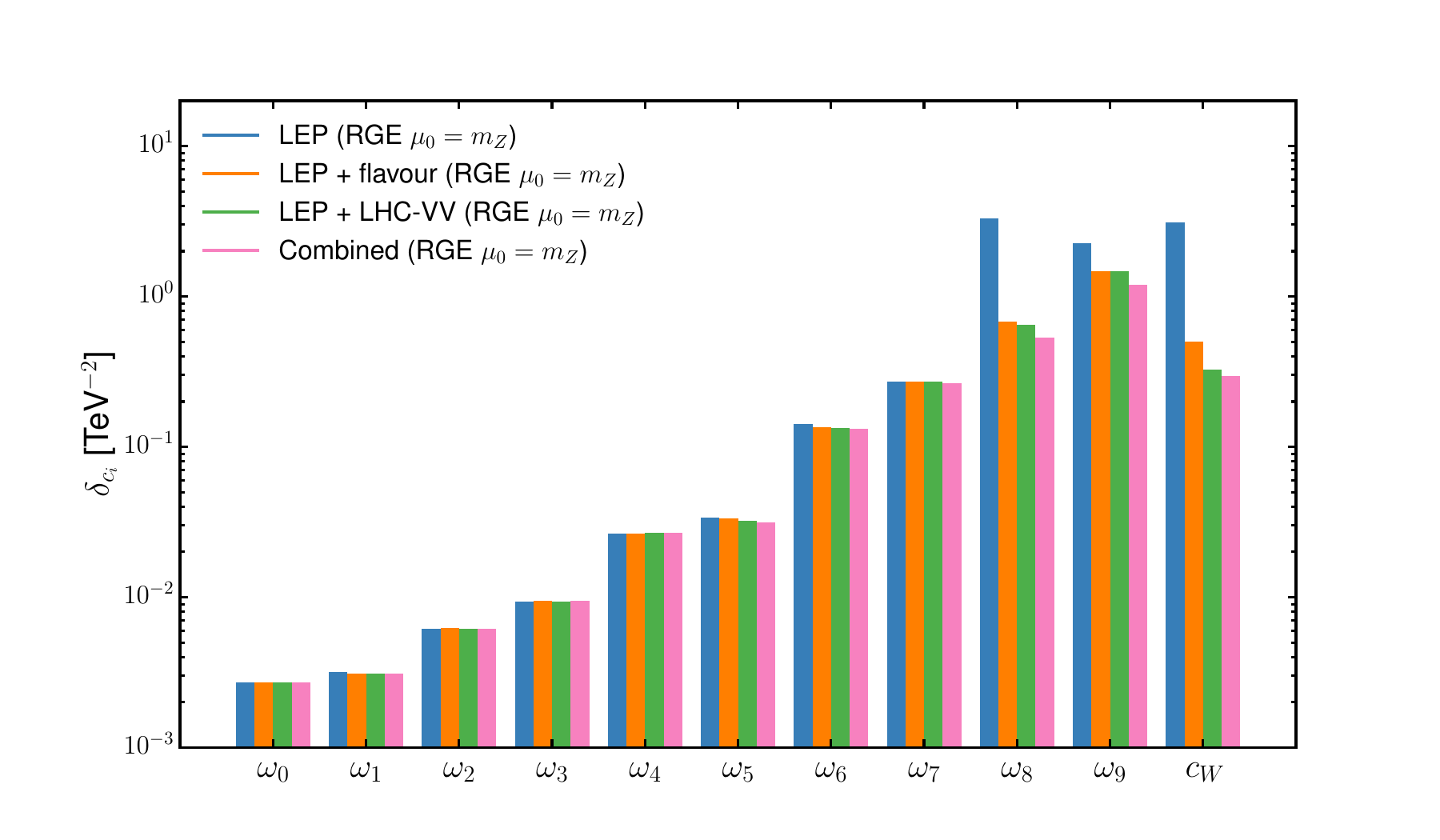}
    \caption{Marginal bounds on the PCA directions defined in App.~\ref{app:pca}, as obtained from different datasets: using only legacy LEP data (blue), adding flavour data (orange), adding LHC diboson data (green), and combining all datasets (pink). The Wilson coefficients are defined at the reference scale $\mu_0 = m_Z$.}
   \label{fig:SMEFT_limits_pca}
\end{figure}

\begin{figure}[t!]
    \centering
      \includegraphics[width=0.45\linewidth, page=10]{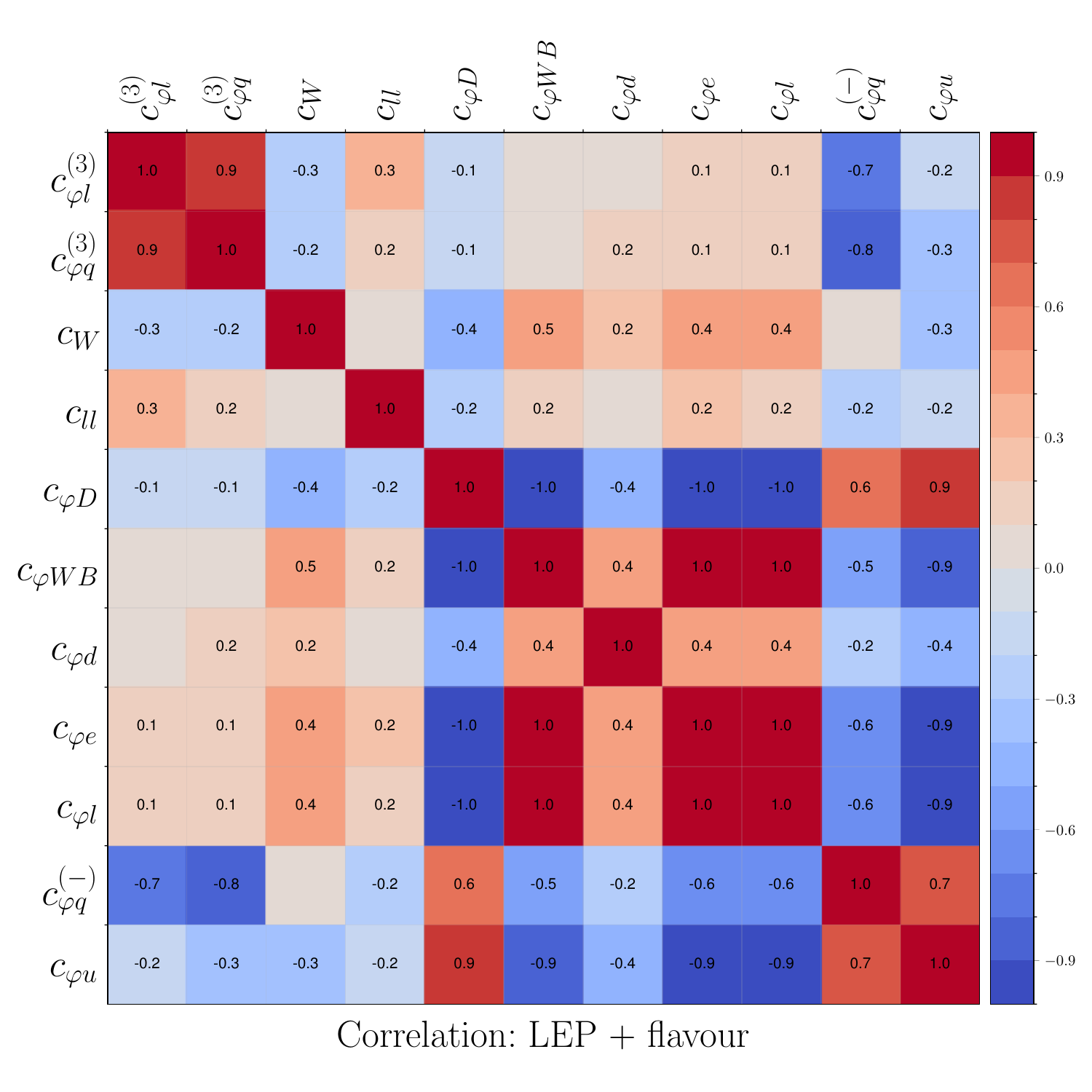}
      \includegraphics[width=0.45\linewidth, page=12]{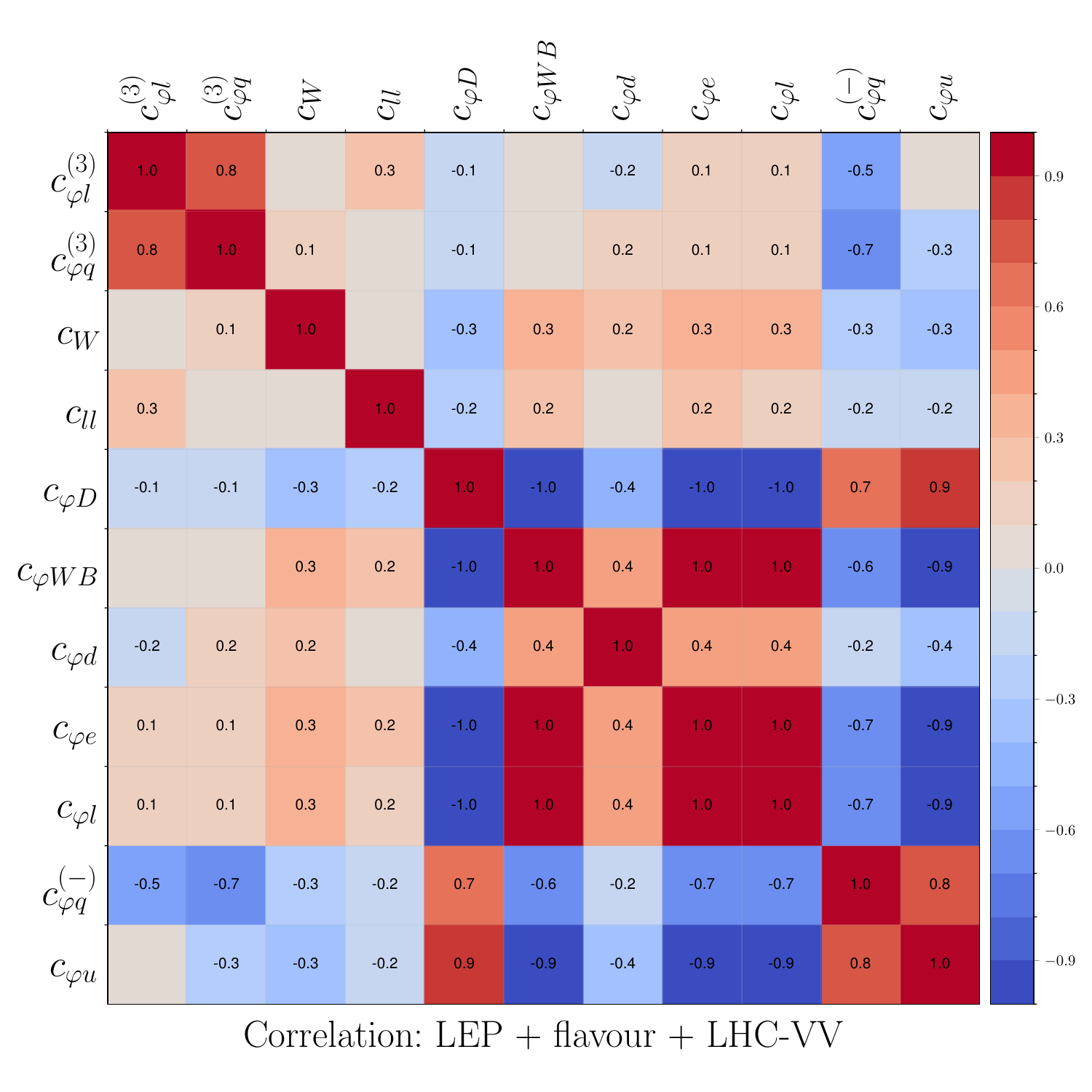}
  \caption{Correlation matrices among the 11 relevant SMEFT operators for different datasets used in the global fit. The left panel shows results using both LEP and flavour data, and the right panel incorporates all available datasets. Wilson coefficients are defined at the reference scale $\mu_0=\Lambda$. The progression highlights how additional data sources affect correlations between operators, reducing degeneracies and improving the robustness of the fit.}
\label{fig:correlation_matrices}
\end{figure}

As in the case of the diboson analysis, we find it convenient to assess the impact of including flavour observables by transforming into the PCA basis defined at the $Z$-pole observables. The bar plot in Figure~\ref{fig:SMEFT_limits_pca} illustrates the marginal standard deviations obtained for all directions under different dataset combinations. The results indicate that the combination of LEP and flavour data yields constraints comparable to those obtained from LEP and LHC diboson data, which only provides a notable improvement in the case of the triple gauge boson operator $\mathcal{O}_W$. Notably, even when including flavour observables, LEP remains the dominant source of constraints on the first seven directions, which correspond to the fermion couplings to the weak bosons. The observed improvement primarily arises from the complementary constraining power associated with differently probed directions in the parameter space.

At the same time, the inclusion of all datasets (LEP, flavour, and LHC diboson) consistently provides the most stringent limits, as the global fit benefits from the complementary constraints across all observables. However, we stress again that while the EFT description is a robust framework for low-energy flavour data, its validity at the higher energies probed by the LHC is not necessarily guaranteed. This caveat underscores the importance of examining alternative datasets to ensure the robustness of the derived limits.

While particularly clean to understand the physics aspects of the analysis, the PCA basis might be quite obscure when it comes to UV matching, even more so in the presence of RGE effects as we are forced to define the Wilson coefficients at the $Z$-pole mass scale. For this reason, we now present the results of the fits in terms of the Wilson coefficients in Tab.~\ref{tab:ops}, defining the reference scale at $1$ TeV.

The correlation matrices in Figure~\ref{fig:correlation_matrices} provide insight into the relationships between the 11 relevant SMEFT operators under different dataset combinations. In the left panel, which considers a combination of legacy LEP data and flavour observables, we observe a significant reduction in correlations between many operators compared to Figure~\ref{fig:correlation_matrices_LEP}. The latter was heavily affected by correlations that degraded the ability to constrain individual operators, as degeneracies limited the discriminating power of the global fit. 
The panel demonstrates that incorporating flavour data dramatically reduces these correlations, enhancing the constraining power of precise LEP measurements.

Finally, the right panel shows that adding LHC diboson data alongside LEP and flavour data does not qualitatively alter the improvement already achieved with flavour data. While LHC data provides independent constraints, its primary role is to further refine the limits rather than significantly modifying the underlying correlation structure.

\begin{figure}[t!]
    \centering
    \includegraphics[width=0.85\linewidth, page=7]{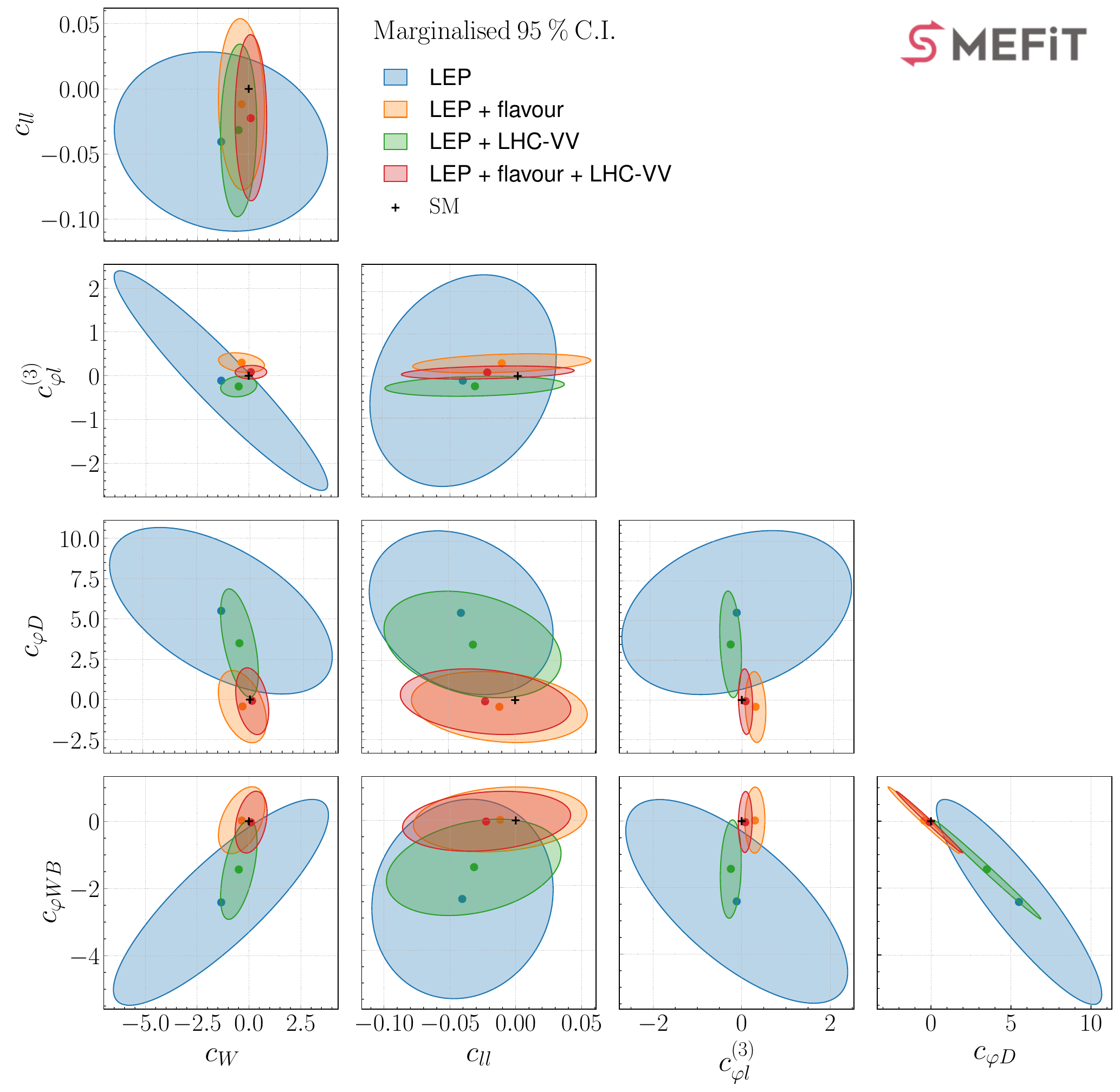}
\caption{Two-dimensional plots showing the marginalized 95\% CL limits on SMEFT coefficients for four different dataset combinations and for a selection of the $11$ Wilson coefficients from Tab.~\ref{tab:ops}. The shaded areas represent the allowed regions for the following cases: using only legacy LEP data (blue), adding flavour data (orange), adding LHC diboson data (green), and including all datasets (red). Wilson coefficients are defined at the reference scale $\mu_0=1$ TeV.}
\label{fig:scatter_matrix}
\end{figure}

To better illustrate the impact of different dataset combinations, Figure~\ref{fig:scatter_matrix} presents the two-dimensional marginalised $95\%$ credible regions of the SMEFT coefficients. For clarity, we select a subset of the 11 operators, focusing on those that contribute to the definition of the TGCs in Eq.~\eqref{eq:deltag}. RGE effects are consistently included in all fits, with the reference scale for each coefficient set at $\mu_0 = 1$ TeV. Each panel displays the allowed parameter space for a pair of SMEFT coefficients, with shaded areas representing the constraints derived from different datasets.

The figure illustrates the progressive tightening of constraints as additional datasets are integrated into the global fit. Notably, incorporating flavour data significantly reduces the allowed parameter space compared to LEP data alone and shifts the constraints toward a more SM-like region, especially when combined with LHC diboson production.

It is important to note that the results presented here are based on a linear fit in the SMEFT coefficients. However, we have also performed quadratic fits to account for higher-order contributions in the SMEFT framework. We find that the qualitative features observed in this analysis remain unchanged in the quadratic fit, indicating the robustness of the results and the conclusions drawn regarding the impact of different datasets on the constraints of SMEFT coefficients. In particular, we find that when all data are combined, the sensitivity to the Wilson coefficients remains largely in the linear regime, with quadratic corrections providing only moderate improvements. The most significant impact is on the triple gauge Wilson coefficient $c_W$, which improves by a factor of 3 when quadratic corrections are included.  However, incorporating the $Zjj$ dataset would diminish this effect, as it already improves the limit on  $c_W$ at linear order, as discussed in~\cite{Ellis:2020unq}. 

Appendix~\ref{app:bounds} presents the collection of numerical results from both the linear and quadratic fits for the Wilson coefficient bounds.

%% file: sec-UV-completions.tex
\section{UV origin of SMEFT coefficients and EFT validity}
\label{sec:uv}
Understanding the possible UV origins of SMEFT coefficients is crucial to evaluating the validity of the description of the EFT in different datasets. The SMEFT framework assumes that the effects of new physics can be captured by a series of higher-dimensional operators, valid only when the energy scales involved are much lower than the scale of new physics $\Lambda$. This assumption is challenged when considering specific datasets, particularly those involving high-energy events.
\begin{figure}[t!]
    \centering
    \includegraphics[width=0.95\linewidth]{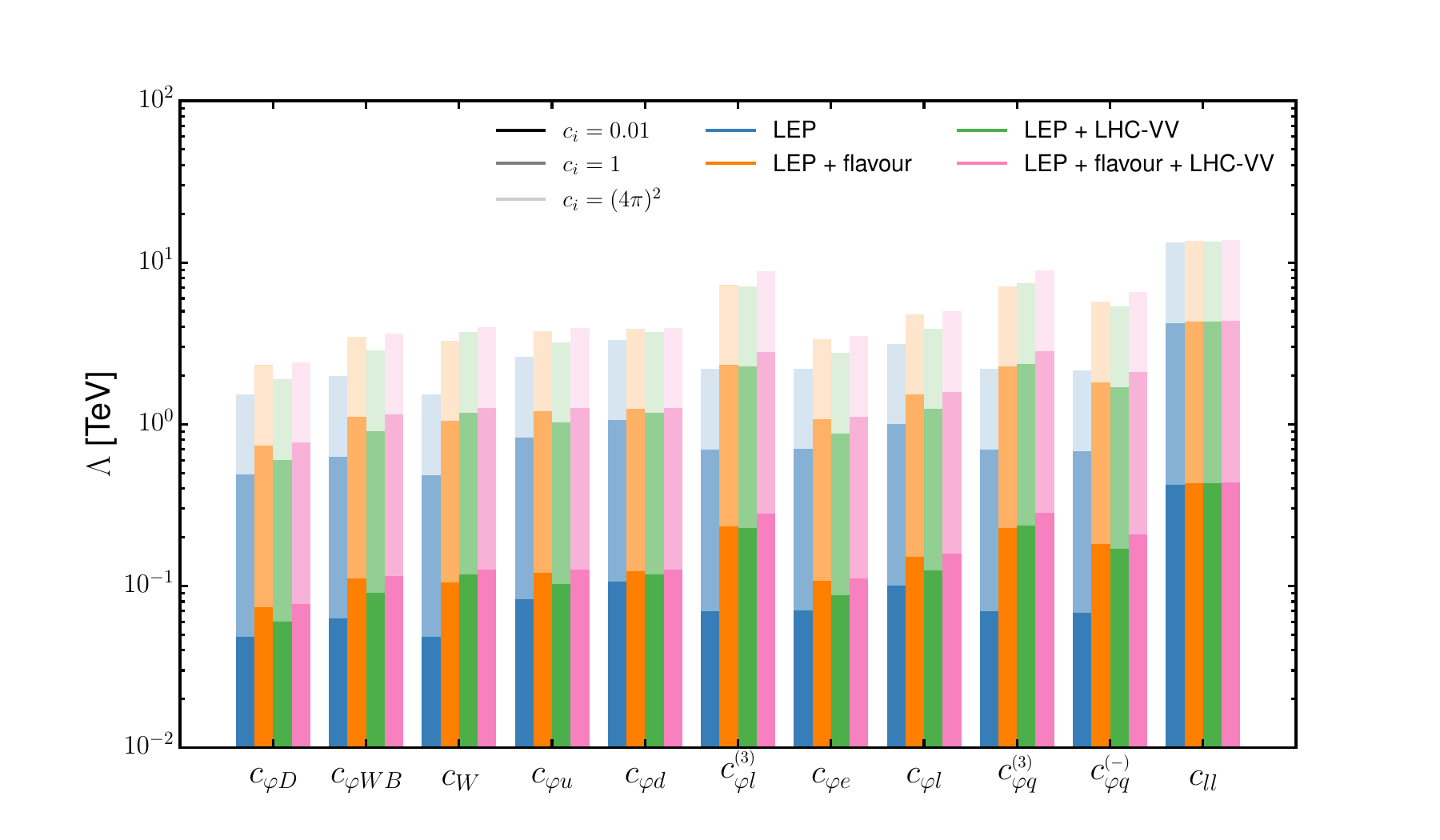}
    \caption{95\% C.L. limits on the  EFT scale associated with each coefficient in a global fit, and for three broad assumptions about the coupling $C_i$= 0.01 (weak coupling, loop suppression), 1 (natural value), and 4$\pi^2$ (strong coupling). The different columns correspond to using the LEP dataset (blue), adding flavour data (orange), adding LHC diboson data (green), combining all together (pink). In all cases fits are performed including RGE effects and defining the Wilson coefficients at the reference scale $\mu_0=1$ TeV.}
    \label{fig:LambdaoverC}
\end{figure}
For example, in the LHC diboson $WW$ channel, the primary variables used to set limits on SMEFT coefficients are sensitive to the energy of the event. A key observable is the transverse momentum ($p_T$) of the lepton from one of the $W$ boson decays. New physics effects are typically searched for in kinematic regions where $p_T$ is high, corresponding to higher center-of-mass energies $\sqrt{\hat s}$. If the underlying UV model includes light intermediate states, parts of this high-$p_T$ kinematic regime may not be accurately described by the EFT approach, as the EFT expansion breaks down when the energy of the event approaches or exceeds the scale $\Lambda$.

For instance, in our fits, we utilise the Run 2 LHC diboson datasets~\cite{ATLAS:2019rob}, selecting the differential distribution $d\sigma/dm_{e\mu}$. Here, the data corresponds to the $e\mu$ decay channel, and $m_{e\mu}$ denotes the invariant mass of the lepton pair. The last bin measured provides information from events with $600~\text{GeV} < m_{e\mu} < 1500~\text{GeV}$. At such high invariant mass for the decay products in the final state, the typical energy of the event is in the TeV range. This should be compared with the results of the global fit, whose interpretation depends on a combination of the mass scale of the new states, $\Lambda$, and their coupling, $C_i$.

To assess the validity of the EFT framework, we compare the experimental kinematic reach ($E = \sqrt{\hat{s}} \gtrsim 1~\text{TeV}$) for the diboson channel with the actual limits obtained from the data. In Fig.~\ref{fig:LambdaoverC}, we present these limits in terms of the mass reach $\Lambda$ as defined in Eq.~\eqref{eq:lambda_def}, which reflects the sensitivity of the experiments to both the mass and coupling of potential new physics. The figure displays limits on $\Lambda$ for various assumptions about the coupling $C_i$, including the simple case of $C_i = 1$, a weakly coupled or one-loop suppressed benchmark of $C_i = 0.01$, and a strongly coupled scenario with extreme coupling $C_i = (4\pi)^2$. The bars represent the limits for different dataset combinations: LEP data only (blue), LEP plus flavour (orange), LEP plus LHC diboson (green), and the combination of all datasets (pink).
Note that to display the results including the effects of RGE, we must choose a reference scale, which in this case is set at $\mu_0 = 1$ TeV. From a bottom-up perspective, this choice is relatively arbitrary; however, its appropriateness depends on the specific UV theory being matched, as different scenarios may favour different reference scales.

Focusing on the LEP plus LHC diboson combination (green bars), we observe that most limits are below $1~\text{TeV}$. This indicates that using the high invariant mass bin for $WW$ production to place constraints on SMEFT coefficients is invalid within the EFT framework. Therefore, for energy scales in this regime, the EFT description would break down unless we focused on strongly coupled scenarios. However, such strongly coupled scenarios face their own theoretical issues related to perturbativity.

If we aim to include perturbative UV theories with states below this energy scale, we should avoid using the high-mass bin in an EFT-based analysis. Instead, a more sophisticated approach is required, as outlined in Ref.~\cite{Lessa:2023tqc}.

\begin{figure}[h!]    
    \centering
    \includegraphics[width=0.5\linewidth]{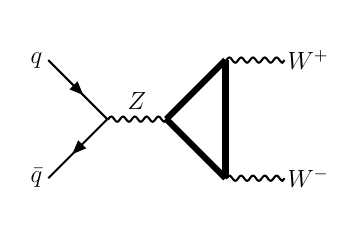}
    \caption{An example of a new physics contribution to $W^+ W^-$ final state with a loop-induced TGC.}
    \label{fig:loopFD}
\end{figure}

For instance, models where new states contribute to TGCs through loops could be considered, as illustrated in Fig.~\ref{fig:loopFD}. In this case, new heavy states would modify the diboson production through a loop-induced effect. In the EFT limit, this effect is suppressed by a loop factor and, at leading order, scales proportionally to $1/M^2$, where $M$ denotes the mass of the heavy state. 
The specific expression of the TGC deviation would depend on the model. For example, in Ref.~\cite{Gorbahn:2015gxa} we computed the matching of the Two Higgs Doublet Models (2HDM). In the decoupling limit, the contribution of those heavy scalar states to the TGCs adopt a simple form. If we focus on the TGC deviations in Eq.~\ref{eq:deltag}, the matching with the 2HDM reads
\begin{equation}
    \delta g_1^Z=- \frac{m_W^2}{96 \pi^2} \frac{2 \tilde \lambda_3+\tilde \lambda_4}{m_{H^+}^2} ,
\end{equation} 
where $\tilde \lambda$ are quartic couplings from the 2HDM potential. Comparing with the relation in Eq.~\ref{eq:deltag} between $\delta g_1^Z$ and the SMEFT coefficients $c_i$, one ends up with a relation,
\begin{equation}
  c_i=\frac{C_i}{\Lambda^2} \simeq \frac{g^2 \tilde \lambda}{96 \pi^2} \frac{1}{m_{H^+}^2 } .
\end{equation}
Looking at this matching, one can identify $C_i$ with the combination of couplings and loop suppression $\frac{g^2 \tilde \lambda}{96 \pi^2}$ which, even for large quartic $\tilde \lambda=4 \pi$, leads to $C_i\sim 10^{-3}$. Comparing this value with the current limits displayed in Fig.~\ref{fig:LambdaoverC}, one would end up placing limits on very light states, outside of the range of validity of the EFT. 

One could think that the issue of loss of validity for the EFT expansion could be avoided if new states would induce TGCs at tree-level instead of at one-loop. In Ref.~\cite{Gorbahn:2015gxa} we proposed such a scenario, where the effects on TGCs come through the mixing of a radion/dilaton state, originated from the breaking of an approximate scale invariant theory at some high scale (dilaton) or, alternatively, the spontaneous compactification of an extra dimension (radion). Those two pictures are related by holography and their connection has been thoroughly studied in the literature, e.g. Ref.~\cite{Hirn:2006wg}.  

In this case, the matching of this theory to the TGCs looks like~\cite{Gorbahn:2015gxa}
\begin{equation}
    \delta g_1^Z = - b_2 \alpha_2 \left(\frac{m_h v}{m_r f}\right)^2 ,
\end{equation}
where $m_{h,r}$ are the SM Higgs and radion masses, $f$ is the scale of conformal breaking (or compactification scale), $v$ is the electroweak vev, $b_2$ is the $SU(2)$ beta function and $\alpha_2=g^2/4 \pi$. The prefactor $b_2 \alpha_2$ receives contributions from the strong sector responsible of the breaking of dilatation symmetry. We can estimate it in the pure conformal field theory (CFT) limit, leading to $b^{CFT}_2 \alpha_2 \simeq 2 \pi/\log (\Lambda_{IR}/\Lambda_{UV})$. In typical warped scenarios~\cite{Csaki:2007ns}, $\log (\Lambda_{IR}/\Lambda_{UV})$ is $\cal O$(30). Translating these estimates into the EFT coefficients we obtain, 
\begin{equation}
  c_i=\frac{C_i}{\Lambda^2} \sim 0.2 \, \frac{m_h^2}{m_r^2} \frac{1}{f^2} ,
\end{equation}
which leads to $C_i \simeq 10^{-3}$ for $m_r$ = 1 TeV and $f=$ 2 TeV, again far away from being tested by current experiments.

These examples should illustrate that the use of an EFT approach for high-energy observables may not be suitable for a perturbative UV completion of the SMEFT. In such cases, instead of using the EFT one should compute a form factor for the TGC that encapsulates the entire kinematic dependence of the dynamical intermediate state, following the methodology described in Ref.~\cite{Lessa:2023tqc}. This form factor, for example, would produce new physics effects in the $m_{e\mu}$ distribution in $W^+W^-$ which would differ from those in a SMEFT description at dimension-six.

On the other hand, flavour observables are measured at much lower energy scales, where the EFT description is robust. Observables such as rare decays, mixing phenomena, and CP violation occur in energy regimes well below the scale of new physics. Consequently, flavour data provides a safe and reliable testing ground for constraining SMEFT coefficients without concerns about the breakdown of the EFT.

To summarise, the UV completion of the SMEFT description thus plays an important role in determining whether the EFT approach is valid for a given dataset. While high-energy datasets such as those from the LHC require careful consideration of their kinematic regimes and potential breakdowns of the EFT, low-energy datasets like flavour observables are inherently well-suited for EFT analyses. This distinction shows the importance of tailoring the interpretation of SMEFT constraints to the specific characteristics of the dataset in question.

%% file: sec-conclusions.tex
\section{Conclusions}
\label{sec:concls}

In this work, we have explored the constraints on flavour-blind SMEFT coefficients through a combination of low-energy flavour observables and high-energy collider datasets. By performing global fits that integrate legacy LEP data, flavour measurements, and LHC diboson data, we demonstrated the complementary strengths of these datasets in probing new physics within the SMEFT framework.

Our analysis shows that flavour observables, measured at low energies, provide precise and robust constraints on SMEFT coefficients, benefiting from the validity of the EFT description in these regimes. The addition of flavour data to LEP datasets significantly reduces degeneracies among SMEFT operators, unlocking the full potential of the high-precision LEP measurements. High-energy LHC data, while offering independent constraints, require careful consideration of its kinematic regimes to ensure the validity of the EFT framework. In particular, we have shown how the current sensitivity to weakly coupled extensions of the SM lies in the range of sub-TeV light intermediate states, which may limit the applicability of EFT analyses in high-energy bins.

The interplay between datasets highlights the importance of combining information from different energy scales. Flavour observables constrain SMEFT coefficients with high confidence in the low-energy regime, while LEP and LHC data probe complementary aspects of the parameter space. This synergy allows us to set stringent bounds on SMEFT operators and provides a more comprehensive picture of potential new physics effects.

Finally, we emphasize that while this study focused on a linear fit in the SMEFT coefficients, a complementary quadratic fit showed no qualitative changes to the results, showing the robustness of our findings. We also examined the impact of RGE and operator mixing across different energy scales, highlighting the non-negligible modifications to Wilson coefficients when transitioning from high-energy collider data to low-energy observables. Our results show the importance of including these effects in global fits to maintain consistency and reliability.

This work could be expanded in several directions. For instance, even if not strictly within the EFT expansion, high-energy diboson data could be included using a form factor description, as advocated in Ref.~\cite{Lessa:2023tqc}, and the model dependence of this form factor could be studied as a source of systematic uncertainty. Additionally, while we focused on the most sensitive rare decays to constrain TGCs, other low-energy observables could be added to the analysis to provide further sensitivity as we move along increasing the precision of the global fits, see for example Ref.~\cite{Bartocci:2023nvp}, where an important interplay is found between EWPOs and parity violation experiments. Finally, although our study found robust results on RGE effects across varying reference scales, a more detailed and systematic investigation of scale dependence within EFT analyses is needed and could potentially enhance the reliability of the conclusions.

%% file: app-pca.tex
\section{Principal Component Analysis of EWPO}
\label{app:pca}

In this appendix, we present the Principal Component Analysis (PCA) of the operators relevant to $Z$ pole observables. In the flavour-blind scenario, ten operators from Table~\ref{tab:ops} can be probed. However, only eight independent directions in the full parameter space can be effectively constrained. 

We report here the definitions of the eigenvectors $\omega_i$ obtained via PCA within the SMEFiT framework. Among these, only the directions $\omega_0, \dots, \omega_7$ can be constrained, whereas $\omega_8$ and $\omega_9$ remain unconstrained. The latter correspond to triple gauge couplings (TGCs) and are associated with the operators $\mathcal{O}_{HB}$ and $\mathcal{O}_{HW}$ in the HISZ basis~\cite{Brivio:2017bnu}, being linear combinations of the $\omega_B$ and $\omega_W$ directions.

\begin{align*}
\omega_0 &= -0.52 \, c^{\scriptscriptstyle(3)}_{\varphi l} +0.41 \, c^{\scriptscriptstyle(3)}_{\varphi q} +0.09 \, c_{ll} -0.24 \, c_{\varphi D} -0.39 \, c_{\varphi W B} -0.04 \, c_{\varphi d} +0.21 \, c_{\varphi e} -0.53 \, c_{\varphi l} \\
&+0.10 \, c_{\varphi q}^{\scriptscriptstyle(-)} +0.05 \, c_{\varphi u} \, ,\\
\omega_1 &= +0.08 \, c^{\scriptscriptstyle(3)}_{\varphi l} +0.36 \, c^{\scriptscriptstyle(3)}_{\varphi q} -0.11 \, c_{ll} +0.41 \, c_{\varphi D} +0.65 \, c_{\varphi W B} -0.03 \, c_{\varphi d} +0.40 \, c_{\varphi e} -0.30 \, c_{\varphi l} \\ 
&+0.08 \, c_{\varphi q}^{\scriptscriptstyle(-)} +0.04 \, c_{\varphi u} \, ,\\
\omega_2 &= -0.31 \, c^{\scriptscriptstyle(3)}_{\varphi l} +0.55 \, c^{\scriptscriptstyle(3)}_{\varphi q} +0.37 \, c_{ll} +0.05 \, c_{\varphi D} +0.17 \, c_{\varphi W B} -0.05 \, c_{\varphi d} -0.27 \, c_{\varphi e} +0.58 \, c_{\varphi l} \\
&+0.15 \, c_{\varphi q}^{\scriptscriptstyle(-)} +0.05 \, c_{\varphi u} \, ,\\
\omega_3 &= +0.08 \, c^{\scriptscriptstyle(3)}_{\varphi l} +0.15 \, c^{\scriptscriptstyle(3)}_{\varphi q} -0.20 \, c_{ll} -0.32 \, c_{\varphi D} +0.35 \, c_{\varphi W B} -0.02 \, c_{\varphi d} -0.74 \, c_{\varphi e} -0.40 \, c_{\varphi l} \\
&+0.04 \, c_{\varphi q}^{\scriptscriptstyle(-)} +0.02 \, c_{\varphi u} \, ,\\
\omega_4 &= +0.27 \, c^{\scriptscriptstyle(3)}_{\varphi l} +0.26 \, c^{\scriptscriptstyle(3)}_{\varphi q} -0.52 \, c_{ll} -0.04 \, c_{\varphi D} -0.26 \, c_{\varphi W B} -0.02 \, c_{\varphi d} +0.05 \, c_{\varphi e} +0.18 \, c_{\varphi l} \\
&+0.67 \, c_{\varphi q}^{\scriptscriptstyle(-)} -0.20 \, c_{\varphi u} \, ,\\
\omega_5 &= +0.45 \, c^{\scriptscriptstyle(3)}_{\varphi l} -0.05 \, c^{\scriptscriptstyle(3)}_{\varphi q} +0.73 \, c_{ll} -0.05 \, c_{\varphi D} -0.03 \, c_{\varphi W B} +0.02 \, c_{\varphi d} -0.25 \, c_{\varphi l} \\
&+0.41 \, c_{\varphi q}^{\scriptscriptstyle(-)} -0.18 \, c_{\varphi u} \, ,\\
\omega_6 &= -0.19 \, c^{\scriptscriptstyle(3)}_{\varphi l} +0.23 \, c_{\varphi D} -0.10 \, c_{\varphi d} -0.12 \, c_{\varphi e} -0.06 \, c_{\varphi l} -0.18 \, c_{\varphi q}^{\scriptscriptstyle(-)} -0.92 \, c_{\varphi u} \, ,\\
\omega_7 &= -0.06 \, c^{\scriptscriptstyle(3)}_{\varphi l} +0.06 \, c^{\scriptscriptstyle(3)}_{\varphi q} +0.12 \, c_{\varphi D} -0.03 \, c_{\varphi W B} +0.98 \, c_{\varphi d} -0.07 \, c_{\varphi e} -0.03 \, c_{\varphi l} \\
&+0.01 \, c_{\varphi q}^{\scriptscriptstyle(-)} -0.05 \, c_{\varphi u} \, ,\\
\omega_8 &= -0.50 \, c^{\scriptscriptstyle(3)}_{\varphi l} -0.50 \, c^{\scriptscriptstyle(3)}_{\varphi q} -0.27 \, c_{\varphi D} +0.40 \, c_{\varphi W B} +0.05 \, c_{\varphi d} +0.14 \, c_{\varphi e} +0.07 \, c_{\varphi l} \\
&+0.48 \, c_{\varphi q}^{\scriptscriptstyle(-)} -0.09 \, c_{\varphi u} \, ,\\
\omega_9 &= +0.23 \, c^{\scriptscriptstyle(3)}_{\varphi l} +0.23 \, c^{\scriptscriptstyle(3)}_{\varphi q} -0.72 \, c_{\varphi D} +0.21 \, c_{\varphi W B} +0.12 \, c_{\varphi d} +0.36 \, c_{\varphi e} +0.18 \, c_{\varphi l} \\
&-0.29 \, c_{\varphi q}^{\scriptscriptstyle(-)} -0.24 \, c_{\varphi u} \, .\\
\end{align*}

%% file: app-bounds.tex
\section{Wilson coefficients numerical bounds}
\label{app:bounds}
In this appendix, we present the numerical values of the  $95\%$  confidence level marginalized intervals for the Wilson coefficients analysed in this study and presented in Sec.~\ref{sec:globalfits}. The results are shown for both linear and quadratic fits, considering different dataset scenarios. Fits are obtained with inclusion of RGE effects and results are reported at the reference scale $\mu_0=1$ TeV. The table provides a direct comparison of the impact of including flavour data and LHC diboson production on the global constraints coming from LEP. These values serve as a reference for the numerical bounds obtained in our analysis.
\begin{table}[h!]
\small
\centering
\begin{tabular}{|c|c|c|c|c|c|}
\hline
 &  & \multicolumn{1}{c|}{All data $\mathcal{O}(1/\Lambda^2)$} & \multicolumn{1}{c|}{All data $\mathcal{O}(1/\Lambda^4)$} & \multicolumn{1}{c|}{LEP + flav $\mathcal{O}(1/\Lambda^2)$} & \multicolumn{1}{c|}{LEP + flav $\mathcal{O}(1/\Lambda^4)$} \\ \hline
Class & Coeffs & 95\% C.I. & 95\% C.I. & 95\% C.I. & 95\% C.I.\\ \hline
\multirow{7}{*}{2FB}
 & $c_{\varphi q}^{(-)}$                                  & [-0.25, 0.2]                                  & [-0.18, 0.14]      & [-0.64, -0.06]             & [-0.65, -0.07] \\ \cline{2-6}
 & $c_{\varphi q}^{(3)}$                    & [-0.1, 0.13]                    & [-0.1, 0.11]                             & [0.08, 0.46]                              & [0.09, 0.46] \\ \cline{2-6}
 & $c_{\varphi u}$ &                 [-0.5, 0.73]             & [-0.35, 0.36]       & [-0.85, 0.5]                   & [-0.85, 0.48] \\ \cline{2-6}
 & $c_{\varphi d}$               & [-1.29, -0.05]                          & [-0.65, 0.1]                   & [-1.17, 0.14]                     & [-0.97, 0.14] \\ \cline{2-6}
 & $c_{\varphi l}$                        & [-0.37, 0.41]                         & [-0.16, 0.34]                       & [-0.33, 0.51]                & [-0.33, 0.51] \\ \cline{2-6}
 & $c_{\varphi l}^{(3)}$                   & [-0.04, 0.21]                    & [-0.07, 0.16]               & [0.12, 0.48]                      & [0.12, 0.48] \\ \cline{2-6}
 & $c_{\varphi e}$                       & [-0.78, 0.8]                   & [-0.36, 0.67]                     & [-0.68, 1.01]                      & [-0.69, 1.01] \\ \cline{2-6}
\hline
\multirow{1}{*}{4l}
 & $c_{ll}$                     & [-0.07, 0.03]                & [-0.07, 0.03]                  & [-0.06, 0.04]                     & [-0.07, 0.04] \\ \cline{2-6}
\hline
\multirow{3}{*}{B}
 & $c_{\varphi WB}$                         & [-0.77, 0.7]                    & [-0.35, 0.6]                      & [-0.8, 0.78]                      & [-0.82, 0.78] \\ \cline{2-6}
 & $c_{\varphi D}$              & [-1.76, 1.6]                & [-1.49, 0.71]                       & [-2.15, 1.42]                           & [-2.14, 1.45] \\ \cline{2-6}
 & $c_{W}$             & [-0.51, 0.69]            & [-0.15, 0.2]                         & [-1.21, 0.57]                          & [-1.34, 0.48] \\ \cline{2-6}
\hline
\end{tabular}
\caption{Marginalized 95\% confidence level intervals for the Wilson coefficients analysed in this study. Results are presented for both linear and quadratic fits, considering two dataset scenarios: ``All data'' includes LEP, flavour, and LHC diboson production, while ``LEP + flav'' excludes diboson production.}
\end{table}

%% file: main.bbl
\providecommand{\href}[2]{#2}\begingroup\raggedright\begin{thebibliography}{10}

\bibitem{Grzadkowski_2010}
B.~Grzadkowski, M.~Iskrzyński, M.~Misiak and J.~Rosiek, \emph{Dimension-six terms in the standard model lagrangian}, \href{https://doi.org/10.1007/jhep10(2010)085}{\emph{Journal of High Energy Physics} {\bfseries 2010} (2010) }.

\bibitem{Brivio_2019}
I.~Brivio and M.~Trott, \emph{The standard model as an effective field theory}, \href{https://doi.org/10.1016/j.physrep.2018.11.002}{\emph{Physics Reports} {\bfseries 793} (2019) 1–98}.

\bibitem{isidori2023standardmodeleffectivefield}
G.~Isidori, F.~Wilsch and D.~Wyler, \emph{The standard model effective field theory at work},  2023.

\bibitem{de_Blas_2020}
J.~de~Blas, D.~Chowdhury, M.~Ciuchini, A.M.~Coutinho, O.~Eberhardt, M.~Fedele et~al., \emph{Hepfit: a code for the combination of indirect and direct constraints on high energy physics models}, \href{https://doi.org/10.1140/epjc/s10052-020-7904-z}{\emph{The European Physical Journal C} {\bfseries 80} (2020) }.

\bibitem{Brivio_2022}
I.~Brivio, S.~Bruggisser, E.~Geoffray, W.~Killian, M.~Krämer, M.~Luchmann et~al., \emph{From models to smeft and back?}, \href{https://doi.org/10.21468/scipostphys.12.1.036}{\emph{SciPost Physics} {\bfseries 12} (2022) }.

\bibitem{Ellis_2021}
J.~Ellis, M.~Madigan, K.~Mimasu, V.~Sanz and T.~You, \emph{Top, higgs, diboson and electroweak fit to the standard model effective field theory}, \href{https://doi.org/10.1007/jhep04(2021)279}{\emph{Journal of High Energy Physics} {\bfseries 2021} (2021) }.

\bibitem{Giani_2023}
T.~Giani, G.~Magni and J.~Rojo, \emph{Smefit: a flexible toolbox for global interpretations of particle physics data with effective field theories}, \href{https://doi.org/10.1140/epjc/s10052-023-11534-7}{\emph{The European Physical Journal C} {\bfseries 83} (2023) }.

\bibitem{Cirigliano:2016nyn}
V.~Cirigliano, W.~Dekens, J.~de~Vries and E.~Mereghetti, \emph{{Constraining the top-Higgs sector of the Standard Model Effective Field Theory}}, \href{https://doi.org/10.1103/PhysRevD.94.034031}{\emph{Phys. Rev. D} {\bfseries 94} (2016) 034031} [\href{https://arxiv.org/abs/1605.04311}{{\ttfamily 1605.04311}}].

\bibitem{Cirigliano:2023nol}
V.~Cirigliano, W.~Dekens, J.~de~Vries, E.~Mereghetti and T.~Tong, \emph{{Anomalies in global SMEFT analyses. A case study of first-row CKM unitarity}}, \href{https://doi.org/10.1007/JHEP03(2024)033}{\emph{JHEP} {\bfseries 03} (2024) 033} [\href{https://arxiv.org/abs/2311.00021}{{\ttfamily 2311.00021}}].

\bibitem{Garosi:2023yxg}
F.~Garosi, D.~Marzocca, A.R.~S\'anchez and A.~Stanzione, \emph{{Indirect constraints on top quark operators from a global SMEFT analysis}}, \href{https://doi.org/10.1007/JHEP12(2023)129}{\emph{JHEP} {\bfseries 12} (2023) 129} [\href{https://arxiv.org/abs/2310.00047}{{\ttfamily 2310.00047}}].

\bibitem{Izaguirre:2016dfi}
E.~Izaguirre, T.~Lin and B.~Shuve, \emph{{Searching for Axionlike Particles in Flavor-Changing Neutral Current Processes}}, \href{https://doi.org/10.1103/PhysRevLett.118.111802}{\emph{Phys. Rev. Lett.} {\bfseries 118} (2017) 111802} [\href{https://arxiv.org/abs/1611.09355}{{\ttfamily 1611.09355}}].

\bibitem{Bauer:2021mvw}
M.~Bauer, M.~Neubert, S.~Renner, M.~Schnubel and A.~Thamm, \emph{{Flavor probes of axion-like particles}}, \href{https://doi.org/10.1007/JHEP09(2022)056}{\emph{JHEP} {\bfseries 09} (2022) 056} [\href{https://arxiv.org/abs/2110.10698}{{\ttfamily 2110.10698}}].

\bibitem{MartinCamalich:2020dfe}
J.~Martin~Camalich, M.~Pospelov, P.N.H.~Vuong, R.~Ziegler and J.~Zupan, \emph{{Quark Flavor Phenomenology of the QCD Axion}}, \href{https://doi.org/10.1103/PhysRevD.102.015023}{\emph{Phys. Rev. D} {\bfseries 102} (2020) 015023} [\href{https://arxiv.org/abs/2002.04623}{{\ttfamily 2002.04623}}].

\bibitem{Gavela:2019wzg}
M.B.~Gavela, R.~Houtz, P.~Quilez, R.~Del~Rey and O.~Sumensari, \emph{{Flavor constraints on electroweak ALP couplings}}, \href{https://doi.org/10.1140/epjc/s10052-019-6889-y}{\emph{Eur. Phys. J. C} {\bfseries 79} (2019) 369} [\href{https://arxiv.org/abs/1901.02031}{{\ttfamily 1901.02031}}].

\bibitem{Albrecht:2019zul}
J.~Albrecht, E.~Stamou, R.~Ziegler and R.~Zwicky, \emph{{Flavoured axions in the tail of B$_{q}$ \textrightarrow{} \ensuremath{\mu}$^{+}$\ensuremath{\mu}$^{-}$ and B \textrightarrow{} \ensuremath{\gamma}$^{*}$ form factors}}, \href{https://doi.org/10.1007/JHEP09(2021)139}{\emph{JHEP} {\bfseries 21} (2020) 139} [\href{https://arxiv.org/abs/1911.05018}{{\ttfamily 1911.05018}}].

\bibitem{Allwicher:2023shc}
L.~Allwicher, C.~Cornella, G.~Isidori and B.A.~Stefanek, \emph{{New physics in the third generation. A comprehensive SMEFT analysis and future prospects}}, \href{https://doi.org/10.1007/JHEP03(2024)049}{\emph{JHEP} {\bfseries 03} (2024) 049} [\href{https://arxiv.org/abs/2311.00020}{{\ttfamily 2311.00020}}].

\bibitem{Aoude:2020dwv}
R.~Aoude, T.~Hurth, S.~Renner and W.~Shepherd, \emph{{The impact of flavour data on global fits of the MFV SMEFT}}, \href{https://doi.org/10.1007/JHEP12(2020)113}{\emph{JHEP} {\bfseries 12} (2020) 113} [\href{https://arxiv.org/abs/2003.05432}{{\ttfamily 2003.05432}}].

\bibitem{Bruggisser:2021duo}
S.~Bruggisser, R.~Sch\"afer, D.~van Dyk and S.~Westhoff, \emph{{The Flavor of UV Physics}}, \href{https://doi.org/10.1007/JHEP05(2021)257}{\emph{JHEP} {\bfseries 05} (2021) 257} [\href{https://arxiv.org/abs/2101.07273}{{\ttfamily 2101.07273}}].

\bibitem{Bruggisser:2022rhb}
S.~Bruggisser, D.~van Dyk and S.~Westhoff, \emph{{Resolving the flavor structure in the MFV-SMEFT}}, \href{https://doi.org/10.1007/JHEP02(2023)225}{\emph{JHEP} {\bfseries 02} (2023) 225} [\href{https://arxiv.org/abs/2212.02532}{{\ttfamily 2212.02532}}].

\bibitem{Bartocci:2023nvp}
R.~Bartocci, A.~Biek\"otter and T.~Hurth, \emph{{A global analysis of the SMEFT under the minimal MFV assumption}}, \href{https://doi.org/10.1007/JHEP05(2024)074}{\emph{JHEP} {\bfseries 05} (2024) 074} [\href{https://arxiv.org/abs/2311.04963}{{\ttfamily 2311.04963}}].

\bibitem{Bellafronte:2023amz}
L.~Bellafronte, S.~Dawson and P.P.~Giardino, \emph{{The importance of flavor in SMEFT Electroweak Precision Fits}}, \href{https://doi.org/10.1007/JHEP05(2023)208}{\emph{JHEP} {\bfseries 05} (2023) 208} [\href{https://arxiv.org/abs/2304.00029}{{\ttfamily 2304.00029}}].

\bibitem{Farina:2016rws}
M.~Farina, G.~Panico, D.~Pappadopulo, J.T.~Ruderman, R.~Torre and A.~Wulzer, \emph{{Energy helps accuracy: electroweak precision tests at hadron colliders}}, \href{https://doi.org/10.1016/j.physletb.2017.06.043}{\emph{Phys. Lett. B} {\bfseries 772} (2017) 210} [\href{https://arxiv.org/abs/1609.08157}{{\ttfamily 1609.08157}}].

\bibitem{Dror:2015nkp}
J.A.~Dror, M.~Farina, E.~Salvioni and J.~Serra, \emph{{Strong tW Scattering at the LHC}}, \href{https://doi.org/10.1007/JHEP01(2016)071}{\emph{JHEP} {\bfseries 01} (2016) 071} [\href{https://arxiv.org/abs/1511.03674}{{\ttfamily 1511.03674}}].

\bibitem{Maltoni:2019aot}
F.~Maltoni, L.~Mantani and K.~Mimasu, \emph{{Top-quark electroweak interactions at high energy}}, \href{https://doi.org/10.1007/JHEP10(2019)004}{\emph{JHEP} {\bfseries 10} (2019) 004} [\href{https://arxiv.org/abs/1904.05637}{{\ttfamily 1904.05637}}].

\bibitem{Jenkins_2013}
E.E.~Jenkins, A.V.~Manohar and M.~Trott, \emph{Renormalization group evolution of the standard model dimension six operators. i: formalism and $\lambda$ dependence}, \href{https://doi.org/10.1007/jhep10(2013)087}{\emph{Journal of High Energy Physics} {\bfseries 2013} (2013) }.

\bibitem{Jenkins:2013sda}
E.E.~Jenkins, A.V.~Manohar and M.~Trott, \emph{{Naive Dimensional Analysis Counting of Gauge Theory Amplitudes and Anomalous Dimensions}}, \href{https://doi.org/10.1016/j.physletb.2013.09.020}{\emph{Phys. Lett. B} {\bfseries 726} (2013) 697} [\href{https://arxiv.org/abs/1309.0819}{{\ttfamily 1309.0819}}].

\bibitem{Jenkins_2014}
E.E.~Jenkins, A.V.~Manohar and M.~Trott, \emph{Renormalization group evolution of the standard model dimension six operators ii: Yukawa dependence}, \href{https://doi.org/10.1007/jhep01(2014)035}{\emph{Journal of High Energy Physics} {\bfseries 2014} (2014) }.

\bibitem{aoude2023renormalisationgroupeffectssmeft}
R.~Aoude, F.~Maltoni, O.~Mattelaer, C.~Severi and E.~Vryonidou, \emph{Renormalisation group effects on smeft interpretations of lhc data},  2023.

\bibitem{Maltoni:2024dpn}
F.~Maltoni, G.~Ventura and E.~Vryonidou, \emph{{Impact of SMEFT renormalisation group running on Higgs production at the LHC}}, \href{https://doi.org/10.1007/JHEP12(2024)183}{\emph{JHEP} {\bfseries 12} (2024) 183} [\href{https://arxiv.org/abs/2406.06670}{{\ttfamily 2406.06670}}].

\bibitem{Bartocci:2024fmm}
R.~Bartocci, A.~Biek\"otter and T.~Hurth, \emph{{Renormalisation group evolution effects on global SMEFT analyses}},  \href{https://arxiv.org/abs/2412.09674}{{\ttfamily 2412.09674}}.

\bibitem{terHoeve:2025gey}
J.~ter Hoeve, L.~Mantani, J.~Rojo, A.N.~Rossia and E.~Vryonidou, \emph{{Connecting Scales: RGE Effects in the SMEFT at the LHC and Future Colliders}},  \href{https://arxiv.org/abs/2502.20453}{{\ttfamily 2502.20453}}.

\bibitem{Celada:2024mcf}
E.~Celada, T.~Giani, J.~ter Hoeve, L.~Mantani, J.~Rojo, A.N.~Rossia et~al., \emph{{Mapping the SMEFT at high-energy colliders: from LEP and the (HL-)LHC to the FCC-ee}}, \href{https://doi.org/10.1007/JHEP09(2024)091}{\emph{JHEP} {\bfseries 09} (2024) 091} [\href{https://arxiv.org/abs/2404.12809}{{\ttfamily 2404.12809}}].

\bibitem{Ellis:2020unq}
J.~Ellis, M.~Madigan, K.~Mimasu, V.~Sanz and T.~You, \emph{{Top, Higgs, Diboson and Electroweak Fit to the Standard Model Effective Field Theory}}, \href{https://doi.org/10.1007/JHEP04(2021)279}{\emph{JHEP} {\bfseries 04} (2021) 279} [\href{https://arxiv.org/abs/2012.02779}{{\ttfamily 2012.02779}}].

\bibitem{Ethier:2021bye}
{\scshape SMEFiT} collaboration, \emph{{Combined SMEFT interpretation of Higgs, diboson, and top quark data from the LHC}}, \href{https://doi.org/10.1007/JHEP11(2021)089}{\emph{JHEP} {\bfseries 11} (2021) 089} [\href{https://arxiv.org/abs/2105.00006}{{\ttfamily 2105.00006}}].

\bibitem{Kassabov:2023hbm}
Z.~Kassabov, M.~Madigan, L.~Mantani, J.~Moore, M.~Morales~Alvarado, J.~Rojo et~al., \emph{{The top quark legacy of the LHC Run II for PDF and SMEFT analyses}}, \href{https://doi.org/10.1007/JHEP05(2023)205}{\emph{JHEP} {\bfseries 05} (2023) 205} [\href{https://arxiv.org/abs/2303.06159}{{\ttfamily 2303.06159}}].

\bibitem{Hurth:2019ula}
T.~Hurth, S.~Renner and W.~Shepherd, \emph{{Matching for FCNC effects in the flavour-symmetric SMEFT}}, \href{https://doi.org/10.1007/JHEP06(2019)029}{\emph{JHEP} {\bfseries 06} (2019) 029} [\href{https://arxiv.org/abs/1903.00500}{{\ttfamily 1903.00500}}].

\bibitem{10.1093/ptep/ptac097}
P.D.~Group, R.L.~Workman, V.D.~Burkert, V.~Crede, E.~Klempt, U.~Thoma et~al., \emph{Review of particle physics}, \href{https://doi.org/10.1093/ptep/ptac097}{\emph{Progress of Theoretical and Experimental Physics} {\bfseries 2022} (2022) 083C01} [\href{https://arxiv.org/abs/https://academic.oup.com/ptep/article-pdf/2022/8/083C01/49175539/ptac097.pdf}{{\ttfamily https://academic.oup.com/ptep/article-pdf/2022/8/083C01/49175539/ptac097.pdf}}].

\bibitem{Brivio:2017bnu}
I.~Brivio and M.~Trott, \emph{{Scheming in the SMEFT... and a reparameterization invariance!}}, \href{https://doi.org/10.1007/JHEP07(2017)148}{\emph{JHEP} {\bfseries 07} (2017) 148} [\href{https://arxiv.org/abs/1701.06424}{{\ttfamily 1701.06424}}].

\bibitem{ALEPH:2005ab}
{\scshape ALEPH, DELPHI, L3, OPAL, SLD, LEP Electroweak Working Group, SLD Electroweak Group, SLD Heavy Flavour Group} collaboration, \emph{{Precision electroweak measurements on the $Z$ resonance}}, \href{https://doi.org/10.1016/j.physrep.2005.12.006}{\emph{Phys. Rept.} {\bfseries 427} (2006) 257} [\href{https://arxiv.org/abs/hep-ex/0509008}{{\ttfamily hep-ex/0509008}}].

\bibitem{ALEPH:2013dgf}
{\scshape ALEPH, DELPHI, L3, OPAL, LEP Electroweak} collaboration, \emph{{Electroweak Measurements in Electron-Positron Collisions at W-Boson-Pair Energies at LEP}}, \href{https://doi.org/10.1016/j.physrep.2013.07.004}{\emph{Phys. Rept.} {\bfseries 532} (2013) 119} [\href{https://arxiv.org/abs/1302.3415}{{\ttfamily 1302.3415}}].

\bibitem{Workman:2022ynf}
{\scshape Particle Data Group} collaboration, \emph{{Review of Particle Physics}}, \href{https://doi.org/10.1093/ptep/ptac097}{\emph{PTEP} {\bfseries 2022} (2022) 083C01}.

\bibitem{Butter_2016}
A.~Butter, O.J.~Éboli, J.~Gonzalez-Fraile, M.~Gonzalez-Garcia, T.~Plehn and M.~Rauch, \emph{The gauge-higgs legacy of the lhc run i}, \href{https://doi.org/10.1007/jhep07(2016)152}{\emph{Journal of High Energy Physics} {\bfseries 2016} (2016) }.

\bibitem{ATLAS:2019rob}
{\scshape ATLAS} collaboration, \emph{{Measurement of fiducial and differential $W^+W^-$ production cross-sections at $\sqrt{s}=13$ TeV with the ATLAS detector}}, \href{https://doi.org/10.1140/epjc/s10052-019-7371-6}{\emph{Eur. Phys. J. C} {\bfseries 79} (2019) 884} [\href{https://arxiv.org/abs/1905.04242}{{\ttfamily 1905.04242}}].

\bibitem{ATLAS:2019bsc}
{\scshape ATLAS} collaboration, \emph{{Measurement of $W^{\pm}Z$ production cross sections and gauge boson polarisation in $pp$ collisions at $\sqrt{s} = 13$ TeV with the ATLAS detector}}, \href{https://doi.org/10.1140/epjc/s10052-019-7027-6}{\emph{Eur. Phys. J. C} {\bfseries 79} (2019) 535} [\href{https://arxiv.org/abs/1902.05759}{{\ttfamily 1902.05759}}].

\bibitem{CMS:2019efc}
{\scshape CMS} collaboration, \emph{{Measurements of the pp $\to$ WZ inclusive and differential production cross section and constraints on charged anomalous triple gauge couplings at $\sqrt{s} =$ 13 TeV}}, \href{https://doi.org/10.1007/JHEP04(2019)122}{\emph{JHEP} {\bfseries 04} (2019) 122} [\href{https://arxiv.org/abs/1901.03428}{{\ttfamily 1901.03428}}].

\bibitem{CMS:2021icx}
{\scshape CMS} collaboration, \emph{{Measurement of the inclusive and differential WZ production cross sections, polarization angles, and triple gauge couplings in pp collisions at $ \sqrt{s} $ = 13 TeV}}, \href{https://doi.org/10.1007/JHEP07(2022)032}{\emph{JHEP} {\bfseries 07} (2022) 032} [\href{https://arxiv.org/abs/2110.11231}{{\ttfamily 2110.11231}}].

\bibitem{Celada:2024cxw}
E.~Celada, G.~Durieux, K.~Mimasu and E.~Vryonidou, \emph{{Triboson production in the SMEFT}}, \href{https://doi.org/10.1007/JHEP12(2024)055}{\emph{JHEP} {\bfseries 12} (2024) 055} [\href{https://arxiv.org/abs/2407.09600}{{\ttfamily 2407.09600}}].

\bibitem{Dawson:2023ebe}
S.~Dawson, D.~Fontes, C.~Quezada-Calonge and J.J.~Sanz-Cillero, \emph{{Matching the 2HDM to the HEFT and the SMEFT: Decoupling and perturbativity}}, \href{https://doi.org/10.1103/PhysRevD.108.055034}{\emph{Phys. Rev. D} {\bfseries 108} (2023) 055034} [\href{https://arxiv.org/abs/2305.07689}{{\ttfamily 2305.07689}}].

\bibitem{Greljo:2023adz}
A.~Greljo and A.~Palavri\'c, \emph{{Leading directions in the SMEFT}}, \href{https://doi.org/10.1007/JHEP09(2023)009}{\emph{JHEP} {\bfseries 09} (2023) 009} [\href{https://arxiv.org/abs/2305.08898}{{\ttfamily 2305.08898}}].

\bibitem{Greljo:2023bdy}
A.~Greljo, A.~Palavri\'c and A.~Smolkovi\v{c}, \emph{{Leading directions in the SMEFT: Renormalization effects}}, \href{https://doi.org/10.1103/PhysRevD.109.075033}{\emph{Phys. Rev. D} {\bfseries 109} (2024) 075033} [\href{https://arxiv.org/abs/2312.09179}{{\ttfamily 2312.09179}}].

\bibitem{Greljo:2023bab}
A.~Greljo, J.~Salko, A.~Smolkovi\v{c} and P.~Stangl, \emph{{SMEFT restrictions on exclusive b \textrightarrow{} u\ensuremath{\ell}\ensuremath{\nu} decays}}, \href{https://doi.org/10.1007/JHEP11(2023)023}{\emph{JHEP} {\bfseries 11} (2023) 023} [\href{https://arxiv.org/abs/2306.09401}{{\ttfamily 2306.09401}}].

\bibitem{Greljo:2022cah}
A.~Greljo, A.~Palavri\'c and A.E.~Thomsen, \emph{{Adding Flavor to the SMEFT}}, \href{https://doi.org/10.1007/JHEP10(2022)005}{\emph{JHEP} {\bfseries 10} (2022) 010} [\href{https://arxiv.org/abs/2203.09561}{{\ttfamily 2203.09561}}].

\bibitem{Greljo:2022jac}
A.~Greljo, J.~Salko, A.~Smolkovi\v{c} and P.~Stangl, \emph{{Rare b decays meet high-mass Drell-Yan}}, \href{https://doi.org/10.1007/JHEP05(2023)087}{\emph{JHEP} {\bfseries 05} (2023) 087} [\href{https://arxiv.org/abs/2212.10497}{{\ttfamily 2212.10497}}].

\bibitem{Matias:2012xw}
J.~Matias, F.~Mescia, M.~Ramon and J.~Virto, \emph{{Complete Anatomy of $\bar{B}_d -> \bar{K}^{* 0} (-> K \pi)l^+l^-$ and its angular distribution}}, \href{https://doi.org/10.1007/JHEP04(2012)104}{\emph{JHEP} {\bfseries 04} (2012) 104} [\href{https://arxiv.org/abs/1202.4266}{{\ttfamily 1202.4266}}].

\bibitem{LHCb:2020lmf}
{\scshape LHCb} collaboration, \emph{{Measurement of $CP$-Averaged Observables in the $B^{0}\rightarrow K^{*0}\mu^{+}\mu^{-}$ Decay}}, \href{https://doi.org/10.1103/PhysRevLett.125.011802}{\emph{Phys. Rev. Lett.} {\bfseries 125} (2020) 011802} [\href{https://arxiv.org/abs/2003.04831}{{\ttfamily 2003.04831}}].

\bibitem{Straub:2018kue}
D.M.~Straub, \emph{{flavio: a Python package for flavour and precision phenomenology in the Standard Model and beyond}},  \href{https://arxiv.org/abs/1810.08132}{{\ttfamily 1810.08132}}.

\bibitem{Bobeth:2015zqa}
C.~Bobeth and U.~Haisch, \emph{{Anomalous triple gauge couplings from $B$-meson and kaon observables}}, \href{https://doi.org/10.1007/JHEP09(2015)018}{\emph{JHEP} {\bfseries 09} (2015) 018} [\href{https://arxiv.org/abs/1503.04829}{{\ttfamily 1503.04829}}].

\bibitem{Lessa:2023tqc}
A.~Lessa and V.~Sanz, \emph{{Going beyond Top EFT}}, \href{https://doi.org/10.1007/JHEP04(2024)107}{\emph{JHEP} {\bfseries 04} (2024) 107} [\href{https://arxiv.org/abs/2312.00670}{{\ttfamily 2312.00670}}].

\bibitem{Gorbahn:2015gxa}
M.~Gorbahn, J.M.~No and V.~Sanz, \emph{{Benchmarks for Higgs Effective Theory: Extended Higgs Sectors}}, \href{https://doi.org/10.1007/JHEP10(2015)036}{\emph{JHEP} {\bfseries 10} (2015) 036} [\href{https://arxiv.org/abs/1502.07352}{{\ttfamily 1502.07352}}].

\bibitem{Hirn:2006wg}
J.~Hirn and V.~Sanz, \emph{{The Fifth dimension as an analogue computer for strong interactions at the LHC}}, \href{https://doi.org/10.1088/1126-6708/2007/03/100}{\emph{JHEP} {\bfseries 03} (2007) 100} [\href{https://arxiv.org/abs/hep-ph/0612239}{{\ttfamily hep-ph/0612239}}].

\bibitem{Csaki:2007ns}
C.~Csaki, J.~Hubisz and S.J.~Lee, \emph{{Radion phenomenology in realistic warped space models}}, \href{https://doi.org/10.1103/PhysRevD.76.125015}{\emph{Phys. Rev. D} {\bfseries 76} (2007) 125015} [\href{https://arxiv.org/abs/0705.3844}{{\ttfamily 0705.3844}}].

\end{thebibliography}\endgroup
